\documentclass[aps,11pt,prd,notitlepage,tightenlines,nofootinbib,showpacs,superscriptaddress]{revtex4-1}

\usepackage{amsmath}
\usepackage{amssymb,amsthm}
\usepackage{mathtools}
\usepackage{mathrsfs}
\usepackage{relsize}
\usepackage{bm}

\usepackage{epsfig,graphics,graphicx,color,xcolor}
\usepackage{soul} %strike out some texts  \st{something}
\usepackage{cancel} %cross out some texts \cancel{sth} \xcancel{sth}
\usepackage{slashed}	%feynman slashed symbols
\usepackage{subfigure}
\usepackage{array}
\usepackage{ragged2e}
\usepackage{lineno}
\usepackage{natbib}
\usepackage{hyperref}
\hypersetup{colorlinks=true,linkcolor=blue,anchorcolor=blue,citecolor=darkgreen, filecolor=blue,urlcolor=blue,bookmarksnumbered=true,
pdfview=FitB
}
\usepackage{enumitem}

\colorlet{darkgreen}{green!50!black}
\colorlet{brightyellow}{yellow!75!red}
\colorlet{orange}{red!50!yellow}
\colorlet{darkgray}{gray!50!black}
\colorlet{darkred}{red!50!black}

\def\dd{{\mathrm{d}}}
\def\imag{{\mathrm{i}}}

\makeatletter
\newcommand*{\transpose}{%
  {\mathpalette\@transpose{}}%
}
\newcommand*{\@transpose}[2]{%
  \raisebox{\depth}{$\m@th#1\intercal$}%
}
\makeatother 

\begin{document}

%opening 
\title{Frame dependence of form factors in light-front dynamics}
\author{Yang~Li}
%\email{yli48@wm.edu}
\affiliation{Department of Physics, College of William \& Mary, Williamsburg, VA 23185}
\affiliation{Department of Physics and Astronomy, Iowa State University, Ames, IA 50011}

\author{Pieter Maris}
\affiliation{Department of Physics and Astronomy, Iowa State University, Ames, IA 50011}

\author{James P. Vary}
\affiliation{Department of Physics and Astronomy, Iowa State University, Ames, IA 50011}

\date{\today}

\begin{abstract}
 In light-front dynamics, form factors are traditionally computed with the ``good current'' $J^+$ within the Drell-Yan frame $q^+=0$. 
Due to truncations imposed in practical calculations, 
 the from factor may acquire frame dependence, which is often neglected. In this work, we explore the form factors in 
 more general frames, preserving the boost covariance. We find the frame dependence of the elastic form factors for mesons is small in basis light-front holography and related models with two-body Fock space truncation. We suggest to use the difference between form factor results from Drell-Yan frame and the ``longitudinal frame''
as a metric for the violation of the Lorentz symmetry due to Fock space truncation. 
\end{abstract}
\maketitle 

\section{Introduction}

In quantum field theory, elastic electromagnetic form factors characterize the structure of a bound state system. They generalize the multipole expansion of charge and current density in nonrelativistic quantum mechanics.
Formally, they are defined as the Lorentz scalars arising in the Lorentz structure decomposition of the hadron matrix element: 
$\langle \psi_h(p+q)|J^\mu(0) |\psi_h(p)\rangle$. 
If the hadron state vector $|\psi_h(p)\rangle$ is known for arbitrary momentum $p$ of interest, the hadron matrix element can be directly obtained. Light-front wave functions (LFWFs) are boost invariant objects hence are particularly advantageous for this task. 
LFWFs are the eigenfunctions of the light-front quantized Hamiltonian operator at a fixed light-front time $x^+ \equiv t + z/c$.
See Refs.~\cite{Perry:1994kp, Zhang:1994ti, Burkardt:1995ct, Harindranath:1996hq, Brodsky:1997de, Carbonell:1998rj, Hiller:2016itl} for 
reviews of this topic. 
 
\begin{figure}
\centering 
\subfigure[\ Parton-number-conserving contributions \label{fig:LFWF_rep_a}]{\includegraphics[width=0.3\textwidth]{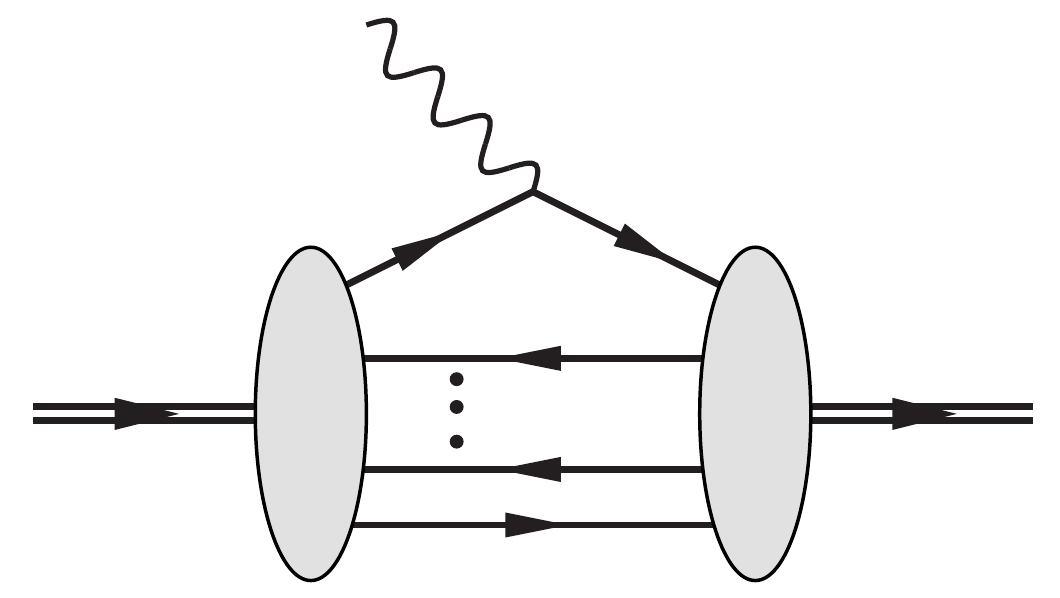}} \qquad
\subfigure[\ Parton-number-non-conserving contributions \label{fig:LFWF_rep_b}]{\includegraphics[width=0.3\textwidth]{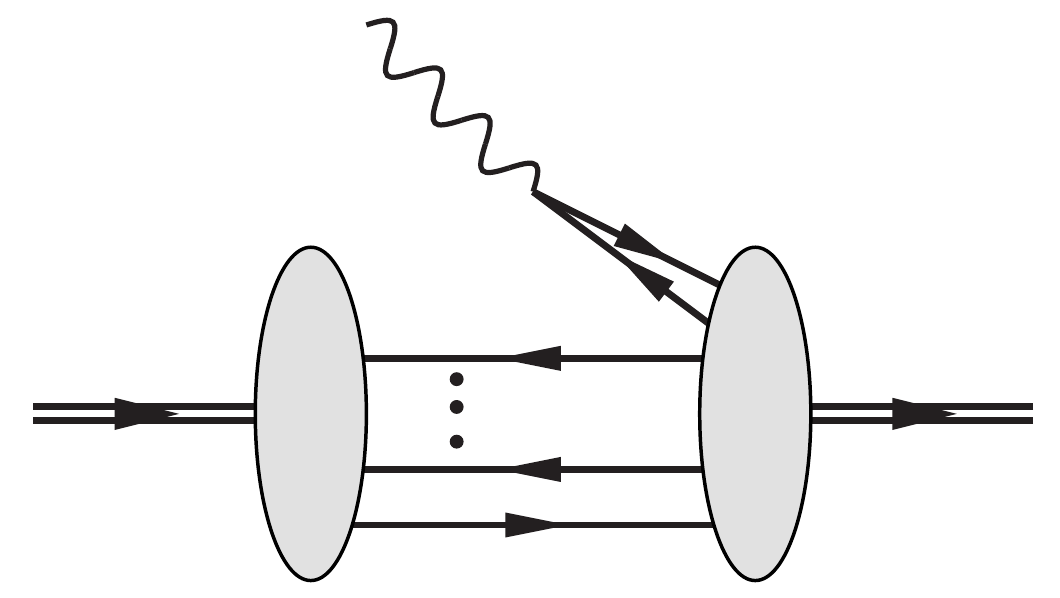}}
\caption{LFWF representation of the hadron matrix element. The double-lines represents the hadrons.
The solid lines represent the partons. The wavy lines represent the probing photon. The shaded areas
represent the LFWFs. These diagrams are ordered by light-front time $x^+$, which flows from left to 
right. 
The parton-number-non-conserving 
contributions (b) stem from pair production/annihilation. }
\label{fig:LFWF_rep}
\end{figure}

Although, by definition, form factors are Lorentz invariants, calculations in light-front dynamics are typically done using 
a specific component $J^+ \equiv J^0 + J^3$, the ``good current'', and in a special frame $q^+=q^0+q^3=0$, the Drell-Yan frame\footnote{Note that it represents infinite
many of frames related by light-front boost transformation.} \cite{Drell:1969km, West:1970av, Brodsky:1973kb, Brodsky:1980zm}. The main advantage of this combination of current and frame choices is that 
vacuum pair production/annihilation is suppressed \cite{Brodsky:1998hn, Brodsky:1973kb, Brodsky:1980zm}. As a result, parton number is conserved and the matrix element only involves the 
overlap of LFWFs of the same parton number [see Fig.~\ref{fig:LFWF_rep_a}]. Nevertheless, this is not an \textit{a priori} restriction and other frames and/or components can be used. However, since the parton number is no longer conserved,  
higher Fock sector wave functions are needed [see Fig.~\ref{fig:LFWF_rep_b}]. 
In most practical calculations, only a finite number of partons can be retained in the Fock space, 
though exceptions exist, e.g. \cite{Chabysheva:2011ed}. Consequently, with truncated Fock sector representations, form factors evaluated in 
different frames or using different components of the current give different results, implying the loss of Lorentz covariance. 

Note that even with the combination
of Drell-Yan frame and $J^+$, higher Fock sector contributions are present. Neglecting these contributions
could also lead to violation of Lorentz symmetry. Since in practical calculations, Fock sector truncation
is part of the model,  the frame dependence of form factors may be used to as a metric of the violation of
Lorentz covariance within the model \cite{Isgur:1988iw}.  This suggestion of a useful metric also applies to other observables. The hope is that, as more Fock sectors are included, the frame dependence may be reduced, 
as is shown in some specific models \cite{Li:2014kfa, Li:2015iaw, Karmanov:2016yzu}.
  
\begin{figure}
\centering 
\subfigure[\  Triangle diagram time ordering a \label{fig:BSA_1}]{\includegraphics[width=0.3\textwidth]{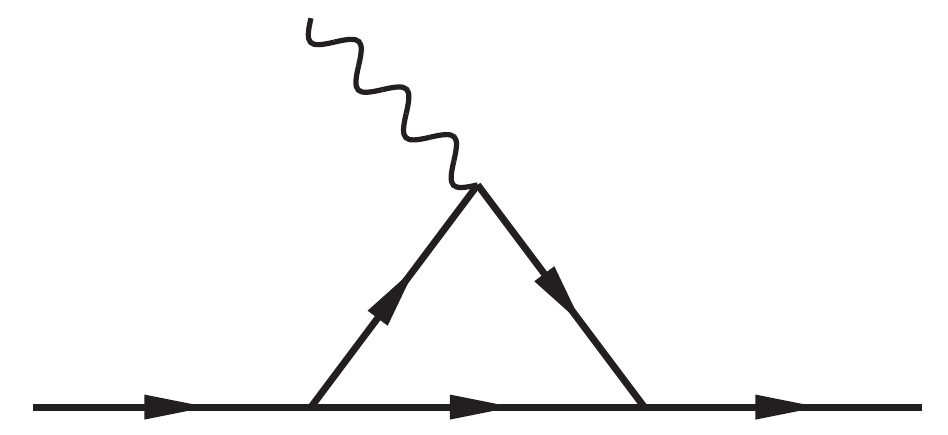}} \qquad
\subfigure[\  Triangle diagram time ordering b \label{fig:BSA_2}]{\includegraphics[width=0.3\textwidth]{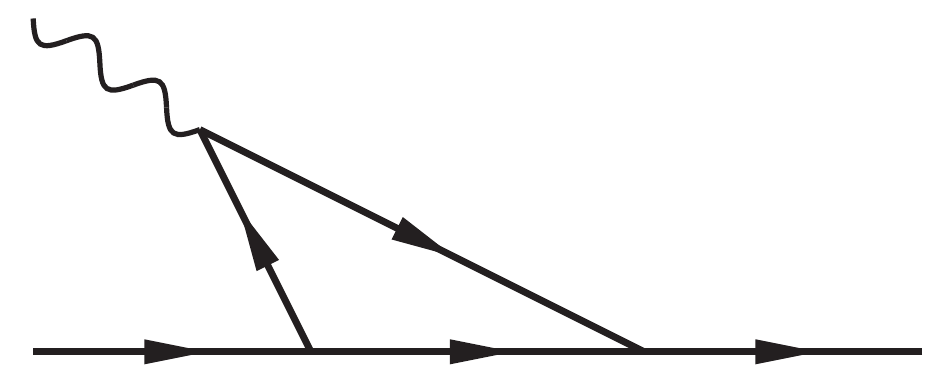}}

\subfigure[\ Overlap contributions \label{fig:BSA_a}]{\includegraphics[width=0.3\textwidth]{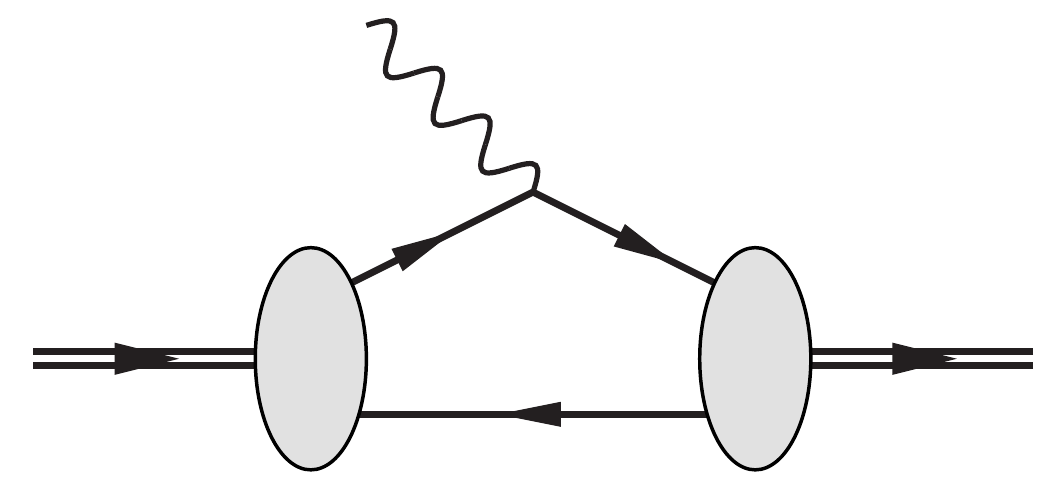}} \qquad
\subfigure[\ Z-diagram contributions \label{fig:BSA_b}]{\includegraphics[width=0.3\textwidth]{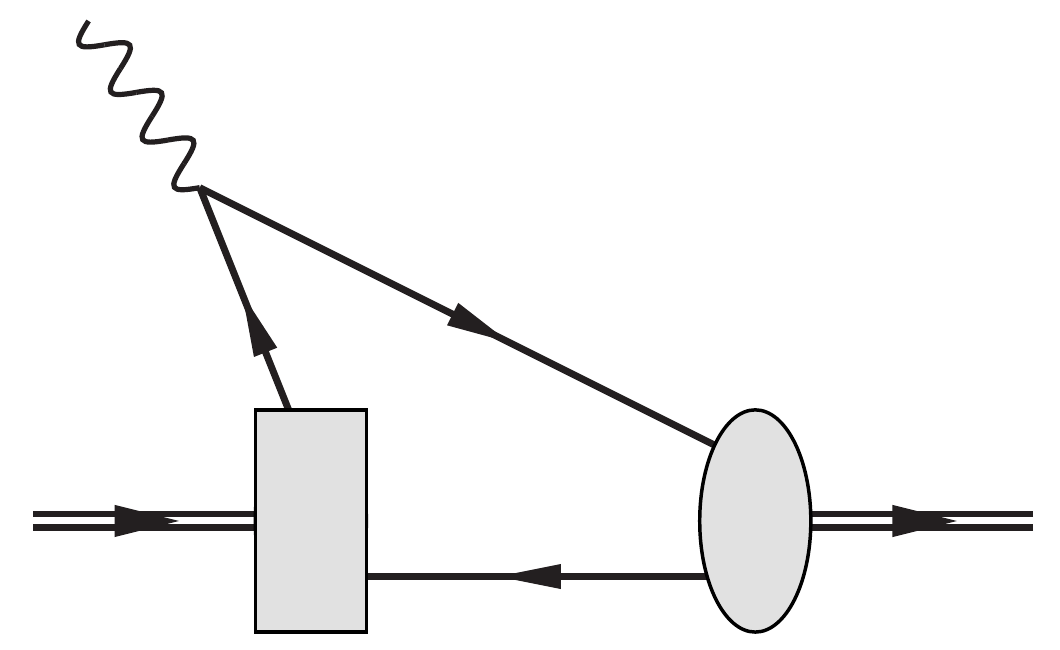}}
\caption{
Top: The light-front projections of the covariant triangle diagrams. (a) and (b) represent different time ordering. 
Bottom: The light-front projection of the Bethe-Salpeter amplitude representation of form factors. 
The shaded oval represents LFWFs. The rectangle represents a non-LFWF vertex. (c) 
resembles (a) while (d) resembles (b). }
\label{fig:BSA}
\end{figure}

The light-front projection of the Bethe-Salpeter amplitude (BSA) provides another insightful perspective. 
In particular, using perturbatively obtained BSA\footnote{For these calculations, one can also start directly
from light-cone perturbation theory, as it is equivalent to the light-front projection of the covariant perturbation theory. }, it was shown that apart from the overlap of LFWFs, a non-LFWF-overlap
contribution, known as the Z-diagrams, also emerges (\cite{Sawicki:1992qj, Brodsky:1998hn, Simula:2002vm, Melikhov:2002mp}, see Fig.~\ref{fig:BSA}). These diagrams are a partial
resummation of higher Fock sector contributions. They do not necessarily vanish even in the Drell-Yan frame except
for $J^+$ for scalar theory \cite{Sawicki:1992qj, Brodsky:1998hn}. 
However, beyond perturbation theory, it is not clear to what extent the Z-diagram is large or 
how to evaluate their contributions, although they can be separately modeled \cite{Choi:1998nf, Mello:2017mor}.  
Furthermore, modern Bethe-Salpeter equations coupled with the Dyson-Schwinger equations are formulated in Euclidean 
space time. To evaluate the form factors, the current may also have to be consistently dressed \cite{Maris:1999bh}. 
The bridge between the Euclidean Bethe-Salpeter and the Minkowskian light-front approaches is not yet built \cite{Gutierrez:2016ixt, dePaula:2016oct, dePaula:2017ikc, Carbonell:2017kqa, Carbonell:2017isq, Kusaka:1995za, Karmanov:2005nv, Sales:1999ec}. 

The dependence of form factors  on the current components and on the reference frame are two typical symptoms  of the violation of the Lorentz covariance in light-front dynamics. 
Model independent analysis of current components in the Drell-Yan frame has been performed extensively by Karmanov and collaborators in the covariant formulation of light-front dynamics \cite{Carbonell:1998rj} (cf.~\cite{Brodsky:1998hn}). Apart from the formal aspects, the use of other current components for certain observables is more
physically justified. For example, in nonrelativistic quantum mechanics, the magnetic moments can only be extracted from 3-dimensional \textit{current density} operator $\vec J$. While in quantum field theory, relativity allows us to extract magnetic moments from the \textit{charge density} operator $J^0$, or in light-front dynamics, $J^+$, this procedure requires a proper implementation of 
 Lorentz covariance within the model. In contrast, the current density operator $\vec J_\perp$, echoing its nonrelativistic 
 counterpart, should provide a more reliable access to magnetic moments, at least in heavy quarkonia \cite{Meijian:2017}.

On the other hand, the investigation of the frame dependence in elastic form factors is rare to the best of our knowledge
\cite{Isgur:1988iw, Sawicki:1992qj, Brodsky:1998hn, Simula:2002vm}. Most studies focus on
the transition form factors in the time-like region where the Drell-Yan frame is not applicable \cite{Cheng:1996if, Choi:1998nf, Brodsky:1998hn, Ji:2000gi}. 
This work is intended to fill the gap. 
In particular, we propose a new parametrization, in which the form factors are expressed as a function of 
two boost invariants $z$, $\Delta_\perp$ [see Eq.~(\ref{eqn:boostinv})]. If the Lorentz covariance is restored, the dependence is reduced to a single Lorentz invariant
$Q^2 = (z^2M^2_h+\Delta^2_\perp)/(1-z)$. 

As a concrete example, we scrutinize the frame dependence of the elastic charge form factor of (pseudo-)scalar mesons for heavy 
quarkonia in a phenomenological model based on light-front holographic QCD.  
In  (pseudo-)scalar mesons, the frame and the component dependence largely separate, so that we can focus on the former.
We discover that the frame dependence is moderate in heavy quarkonia. We also find the frame dependence can be  characterized by 
the discrepancy between two frames: the Drell-Yan frame and the ``longitudinal frame'', the meaning of which
will be explained later. Finally, we will also comment on the frame dependence in light meson elastic form factors
based on calculations with light-front holographic QCD wave functions. 
  
This paper is organized as follows. In Sect.~\ref{sec:formulation} we present the formalism for 
computing form factors in a general frame. In Sect.~\ref{sec:application} we apply the formulation to 
light-front wave functions for heavy quarkonia. 
We conclude in Sect.~\ref{sec:conclusion}.
  
\section{Form factor in light-front dynamics}\label{sec:formulation}

The Lorentz decomposition of the matrix element of a (pseudo)scalar meson $h$ is \cite{Carbonell:1998rj},
\begin{equation}\label{eqn:ff}
\Gamma^\mu(p, p'; \omega) \equiv \langle \psi_h(p'; \omega)|J^\mu(0)|\psi_h(p; \omega)\rangle = (p+p')^\mu F(z, Q^2) + \frac{\omega^\mu}{\omega\cdot p'} S(z, Q^2), 
\end{equation}
where $q_\mu = p'_\mu-p_\mu$, $z = \omega\cdot q/\omega\cdot p'$, $Q^2 = -q^2$, $\omega$ is a \textit{fixed} null 4-vector
($\omega_\mu\omega^\mu = 0$) indicating the orientation of the quantization surface. For the elastic form factor, 
$p^2=p'^2=M^2_h$. Because the light-front quantized state vector $|\psi_h\rangle$ depends on $\omega$,\footnote{The dependence is always in the form of a ratio, e.g. $\omega^\mu/\omega\cdot p$, as there is an additional conformal symmetry, i.e. $\lambda \omega^\mu$ 
and $\omega^\mu$ indicate the same quantization surface \cite{Carbonell:1998rj}.} in covariant light-front dynamics (CLFD, \cite{Carbonell:1998rj}), the Lorentz structure of the hadronic matrix element is extended. Similar analysis can be applied to the decay constant as well as the non-local matrix elements, e.g. distribution amplitude and generalized parton distribution. In the case of (pseudo)scalar mesons, apart from the physical  form factor $F$, there appears a spurious form factor $S$. Furthermore, the form factors depend on two Lorentz scalars $Q^2$ and $z$.\footnote{$z$ is not a Lorentz invariant, as $\omega$ is a fixed 4-vector. } If the Lorentz symmetry is \textit{dynamically} restored,  
the spurious form factor $S$ will vanish, as will the dependence of the form factors on $z$. 

The emergence of the spurious form factors in CLFD is general.  
For spin-1/2 hadrons, there are 3 spurious form factors; for spin-1, the number is 8 \cite{Carbonell:1998rj}.
These spurious form factors are expected to vanish dynamically when the truncations are lifted. For an \textit{ab initio} calculation, our best hope is
that these form factors are suppressed by powers of $\Lambda_\textsc{qcd}/\Lambda$, where $\Lambda$ is the UV scale associated with the truncation. 

In this work, we choose the standard light-front dynamics, 
i.e. $\omega = (\omega^0, \vec \omega) = (1, 0, 0, -1)$. The light-cone coordinates are 
defined as $v^\pm = v^0\pm v^3$, and $\vec v_\perp = (v^1, v^2)$. In this convention, $\omega^- = 2, \omega^+ = \omega_\perp = 0$, and $\omega\cdot v = v^+$. 
Note that this choice does not automatically make the spurious form factor $S$ or the frame dependence disappear. 
Indeed, the spurious form factor $S$ enters the hadron matrix elements of $J^-$, and leads to the violation of the current conservation:
\begin{equation}
q_\mu \Gamma^\mu(p, p'; \omega) = z S(z, Q^2).
\end{equation}  
The physical form factor $F(z, Q^2)$ can be extracted from either $J^+$ or $\vec J_\perp$:
\begin{align}
\langle \psi_h(p'; \omega)|J^+(0)|\psi_h(p; \omega)\rangle =\,& (p^++p'^+) F(z, Q^2), \label{eqn:Jplus} \\
\langle \psi_h(p'; \omega)|\vec J_\perp(0)|\psi_h(p; \omega)\rangle =\,& (\vec p_\perp+\vec p'_\perp) F(z, Q^2). \label{eqn:Jperp}
\end{align}
The current components $J^+$ and $\vec J_\perp$ are related by a kinematical
boost in the transverse direction:
\begin{multline}
\langle \psi_h(p'^+, \vec p'_\perp+p'^+\vec\beta_\perp; \omega)| \vec J_\perp | \psi_h(p^+, \vec p_\perp+p^+\vec \beta_\perp; \omega)\rangle
= \\
\langle \psi_h(p'^+, \vec p'_\perp; \omega)| \vec J_\perp | \psi_h(p^+, \vec p_\perp; \omega)\rangle 
+
\vec\beta_\perp \langle \psi_h(p'^+, \vec p'_\perp; \omega)| J^+ | \psi_h(p^+, \vec p_\perp; \omega)\rangle.
\end{multline}
Substituting  (\ref{eqn:Jplus}, \ref{eqn:Jperp}), these two current components lead to the same results for the form factor $F(z, Q^2)$, as expected.

Next, we turn to the frame dependence, i.e. $z$ dependence of the physical charge form factor $F(z, Q^2)$.  
The meson state vector $|\psi_h(p, j, \lambda)\rangle$ can be expanded in the Fock space,
\begin{equation}
\begin{split}
|\psi_h(p, j, \lambda)\rangle 
 =\,& \sum_{n} \prod_{i=1}^n\sum_{s_i} \int \frac{\dd x_i}{2x_i} \frac{\dd^2k_{i\perp}}{(2\pi)^3} 2\delta(x_1+x_2+\cdots +x_n)(2\pi)^3\delta^2(\vec k_{1\perp}+\vec k_{2\perp}+\cdots+\vec k_{n\perp})\\
 &\times \psi_{h/n}(\{x_i, \vec k_{i\perp}, s_i\}) \, c^\dagger_{s_1}(x_1p^+, \vec k_{1\perp}+x_1\vec p_\perp) \times \cdots \times c^\dagger_{s_n}(x_np^+, \vec k_{n\perp}+x_n\vec p_\perp) |0\rangle \\
 =\,& \sum_{s, \bar s} \int_0^1\frac{\dd x}{2x(1-x)} \int\frac{\dd^2k_\perp}{(2\pi)^3} \psi^{(\lambda)}_{s\bar s/h}(x, \vec k_\perp) \\
 & \times b^\dagger_s\big(xp^+, \vec k_\perp+x\vec p_\perp\big) d^\dagger_{\bar s}\big((1-x)p^+, -\vec k_\perp+(1-x)\vec p_\perp\big) |0\rangle + \cdots 
\end{split}
\end{equation}
where $\psi_{h/n}(\{x_i, \vec k_{i\perp}, s_i\})$ are the LFWFs. The dots 
represent the higher Fock sector contributions. 
The current operator $J^\mu = \overline\psi \gamma^\mu \psi$, where the quark field operator $\psi$ at $x^+=0$ is,
\begin{equation}
\psi(x) = \sum_s \int\frac{\dd p^+\dd^2p_\perp}{(2\pi)^32p^+} \vartheta(p^+) \Big[ b_s(p^+, \vec p_\perp) u_s(p^+, \vec p_\perp) e^{-\imag p\cdot x}
+ d^\dagger_s(p^+, \vec p_\perp) v_s(p^+, \vec p_\perp)  e^{+\imag p\cdot x} \Big]\Big|_{x^+=0}.
\end{equation}
The operators $b$ and $d$ satisfy the standard anticomutation relation,
\begin{equation}
\Big\{b_s(p^+, \vec p_\perp),  b^\dagger_{s'}(p'^+, \vec p'_\perp)\Big\} 
= 
\Big\{d_s(p^+, \vec p_\perp),  d^\dagger_{s'}(p'^+, \vec p'_\perp)\Big\} 
=
(2\pi)^3 2p^+ \delta(p^+-p'^+)\delta^2(\vec p_\perp-\vec p'_\perp) \delta_{ss'}.
\end{equation}
Then the LFWF representation of the electromagnetic vertex is, 
\begin{multline}
\Gamma^\mu_{\lambda'\lambda}(p, p') = \sum_{s, \bar s}\int_0^1 \frac{\dd x}{2x(1-x)} \int\frac{\dd^2 k_\perp}{(2\pi)^3} 
\frac{1}{x'} \sum_{s'} \bar u_{s'}(x'p'^+, \vec k'_\perp+x'\vec p'_\perp) \gamma^\mu u_s(xp^+, \vec k_\perp+x\vec p_\perp) \\
\times \psi^{(\lambda')*}_{s'\bar s/h}(x', \vec k'_\perp) \psi^{(\lambda)}_{s\bar s/h}(x, \vec k_\perp) + \cdots,
\end{multline}
where $x' p'^+ = xp^+ + q^+$, and $\vec k'_\perp+x'\vec p'_\perp = \vec k_\perp + x\vec p_\perp + \vec q_\perp$. The dots 
represent the higher Fock sector contributions. Here for mesons with quark and antiquark having the same flavor, we have coupled the photon only to the quark. Otherwise, the form 
factor is exactly zero due to charge conjugation symmetry.

We introduce a boost invariant\footnote{Note that $\vec q_\perp$, in general, is not a boost invariant.}: 
\begin{equation} \label{eqn:boostinv}
\vec \Delta_\perp = \vec q_\perp - z\vec p'_\perp = p^+ \Big(\frac{\vec p'_\perp}{p'^+} - \frac{\vec p_\perp}{p^+}\Big),
\end{equation} 
where $z = q^+ / p'^+$ is another boost invariant introduced above [see Eq.~(\ref{eqn:ff})].
Using $z$ and $\Delta_\perp$, 
\begin{equation}
x' = x + z(1-x), \quad \vec k'_\perp = \vec k_\perp + (1-x)\vec\Delta_\perp.
\end{equation} 
The momentum fraction in the valence LFWF is constrained by $0 \le x \le 1$. Therefore, we only have  
access to the kinematical region $0\le z < 1$. Negative $z$ probes sea quark contributions in higher Fock sectors. $z>1$ probes 
the time-like region, which is a different process in light-front dynamics. 
The Lorentz invariant momentum transfer squared $q^2 = (p'-p)^2$ depends on these two boost invariants,
\begin{equation}
q^2  = - \frac{z^2M^2_h + \Delta^2_\perp}{1-z} \equiv -Q^2, 
\end{equation}
where $M_h$ is the mass eigenvalue of the meson, i.e. $p^2 = p'^2 = M^2_h$. For the available kinematic range ($0 \le z < 1$), $q^2
\le 0$, i.e. $q^2$ is space-like. 
We introduce two special frames:
\begin{enumerate}
\item[(I)] transverse frame ($z=0$), also known as Drell-Yan frame $(q^+=0)$: $q^2 = -\Delta_\perp^2 = - q_\perp^2$; 
\item[(II)] longitudinal frame $(\Delta_\perp=0)$: $q^2 = - z^2M^2_h/(1-z)$.
\end{enumerate}
Our definition of the longitudinal frame is very similar to the longitudinal frame ($\vec q_\perp=0$) introduced in the literature
\cite{Isgur:1988iw, Sawicki:1992qj, Brodsky:1998hn, Simula:2002vm}. However, $\vec\Delta_\perp$ is boost invariant while $\vec q_\perp$ is not. 
As we have mentioned earlier, the form factor in light-front dynamics depends on two boost invariants $z$ and $\Delta_\perp$, 
instead of one Lorentz invariant $Q^2$. This dependence is referred to as the frame dependence. Note that each pair of
$(z, \Delta_\perp)$ denotes infinitely many reference frames related by light-front boost, longitudinal and transverse, as well as by rotation
in the transverse plane. 

 The LFWF representation of the charge form factor is,
\begin{multline}
F(z, Q^2) = \frac{\sqrt{1-z}}{1-\frac{1}{2}z}
 \sum_{s, \bar s}\int_0^1 \frac{\dd x}{2x(1-x)} \int\frac{\dd^2 k_\perp}{(2\pi)^3} 
 \sqrt{\frac{x}{x+z(1-x)}} \\
\times \psi^{*}_{s\bar s/h}\big( x+z(1-x), \vec k_\perp+(1-x)\vec\Delta_\perp \big) \psi_{s\bar s/h}\big( x, \vec k_\perp \big)
+ \cdots
\end{multline}
where $Q^2 = (z^2M^2_h+\Delta^2_\perp)/(1-z)$. At $Q\to0$, $z\to 0$ and $\Delta_\perp\to0$ and $F(z, Q^2) \to 1$. 
At large $Q$, either large $\Delta_\perp$ or $z\to1$, $F(z, Q^2) \to 0$. 
The explicit expression including higher Fock sectors is presented in Appendix \ref{sec:app}.
In the Drell-Yan frame ($z=0$, $Q^2 = \Delta^2_\perp$), we obtain the familiar expression
\begin{equation}
F_\textsc{dy}(Q^2) = 
 \sum_{s, \bar s}\int_0^1 \frac{\dd x}{2x(1-x)} \int\frac{\dd^2 k_\perp}{(2\pi)^3} 
 \psi^{*}_{s\bar s/h}\big(x, \vec k_\perp+(1-x)\vec\Delta_\perp \big) \psi_{s\bar s/h}\big( x, \vec k_\perp \big)
+ \cdots
\end{equation}
In the longitudinal frame ($\Delta_\perp = 0$, $Q^2 = z^2M^2_h/(1-z)$), 
\begin{multline}
F_\text{long}(Q^2) = 
\frac{\sqrt{1-z}}{1-\frac{1}{2}z}
 \sum_{s, \bar s}\int_0^1 \frac{\dd x}{2x(1-x)} \int\frac{\dd^2 k_\perp}{(2\pi)^3} 
 \sqrt{\frac{x}{x+z(1-x)}} \\
\times \psi^{*}_{s\bar s/h}\big( x+z(1-x), \vec k_\perp\big) \psi_{s\bar s/h}\big( x, \vec k_\perp \big)
+ \cdots
\end{multline}

\section{Application to heavy quarkonia}\label{sec:application}

Recently, we proposed a model for heavy quarkonia based on light-front holographic QCD \cite{Brodsky:2014yha} and one-gluon 
exchange \cite{Wiecki:2014ola}. The theory is solved in the basis function approach (BLFQ, \cite{Li:2015zda, Li:2017mlw}). 
The one-gluon exchange implements the short-distance physics and supplies the proper spin structure. 
The resulting LFWFs have been used to compute a number of observables as well as in diffractive vector
meson production, showing reasonable agreement with the available experimental data \cite{Chen:2016dlk, Vary:2016emi}. 
In this model, the violation of Lorentz symmetry leads to the the spread of mass eigenvalues with the same angular momentum 
$j$ but different magnetic projection $m_j$. However, such violation is sufficiently small that it does not interfere with spectrum reconstruction. 
Therefore, it is interesting to see whether the frame dependence,  also originating from the violation of the rotational symmetry, 
is under control.   

As mentioned, the model is solved in a basis function approach. The LFWFs read, 
\begin{equation}
\psi_{s\bar s}(x, \vec k_\perp) = \sum_{n, m, l} \psi(n, m, l, s, \bar s) \phi_{nm}(\vec k_\perp/\sqrt{x(1-x)}) 
X_l(x).
\end{equation}
Here $\phi_{nm}$ and $X_l$ are known analytic functions (see Refs.~\cite{Li:2015zda, Li:2017mlw} for details). 
The basis space is truncated by $2n+|m|+1\le N_{\max}$, and $l \le L_{\max}$, and in the calculation, 
$N_{\max} = L_{\max}$ is chosen. The truncation introduces a UV scale $\Lambda_\textsc{uv} \approx
\kappa \sqrt{N_{\max}}$ and a resolution in the longitudinal direction $\Delta x \approx L_{\max}^{-1}$, 
where $\kappa$ is the confining strength whose value is given in Ref.~\cite{Li:2017mlw}. Since form factors 
in light-front dynamics are represented as the convolution of LFWFs, the variables $z$, $\Delta_{\perp}$ and $Q^2$ are   
only supported up to the basis resolutions: $Q^2 \lesssim \kappa^2 N_{\max}$
in the transverse direction and $Q^2 \lesssim M_h^2 {L_{\max}}$ in the longitudinal direction.  Beyond these regimes, the 
LFWFs are dominated by the asymptotics of the basis included within the limited basis space.

\begin{figure}
%\centering
\includegraphics[width=0.48\textwidth]{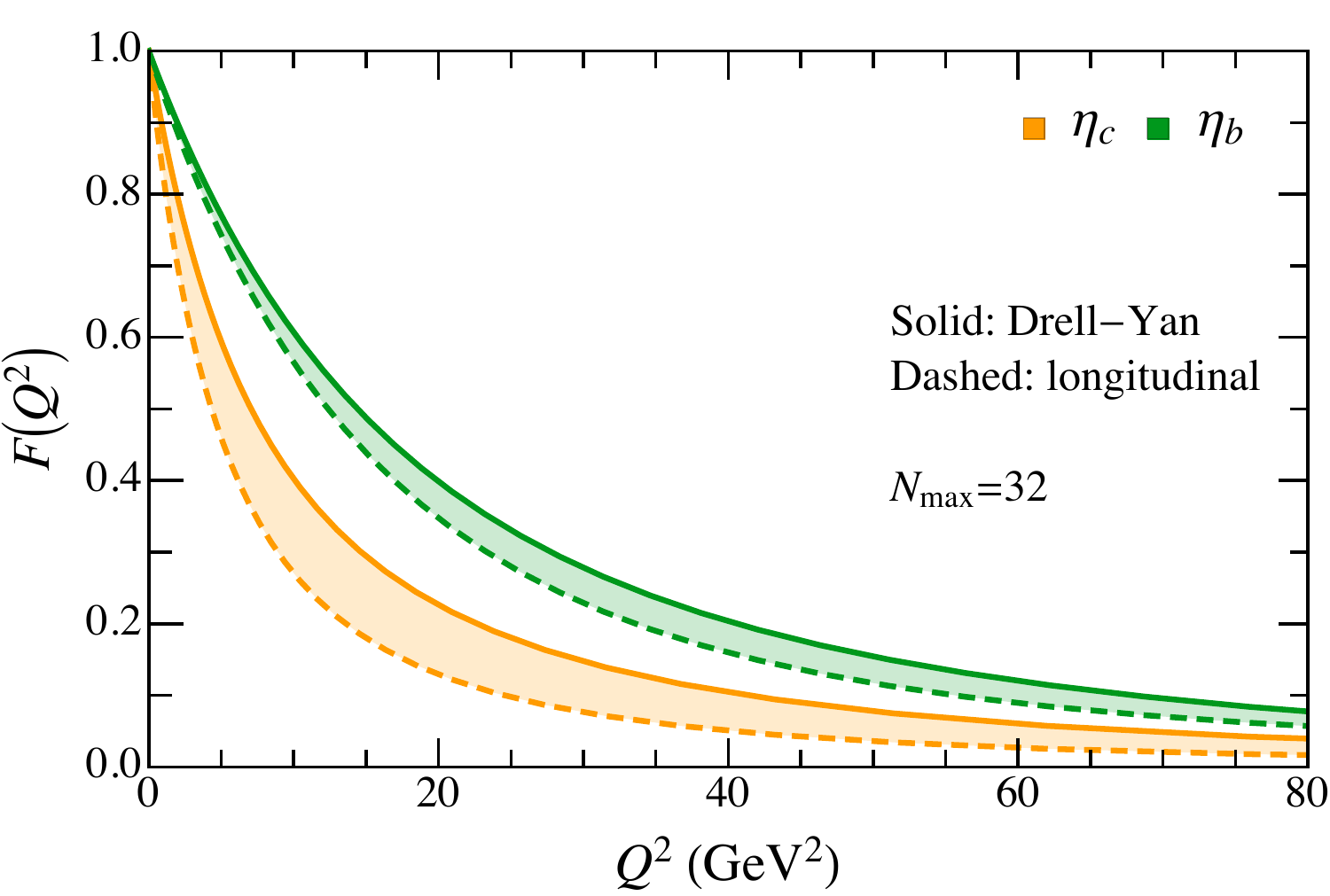}
\includegraphics[width=0.48\textwidth]{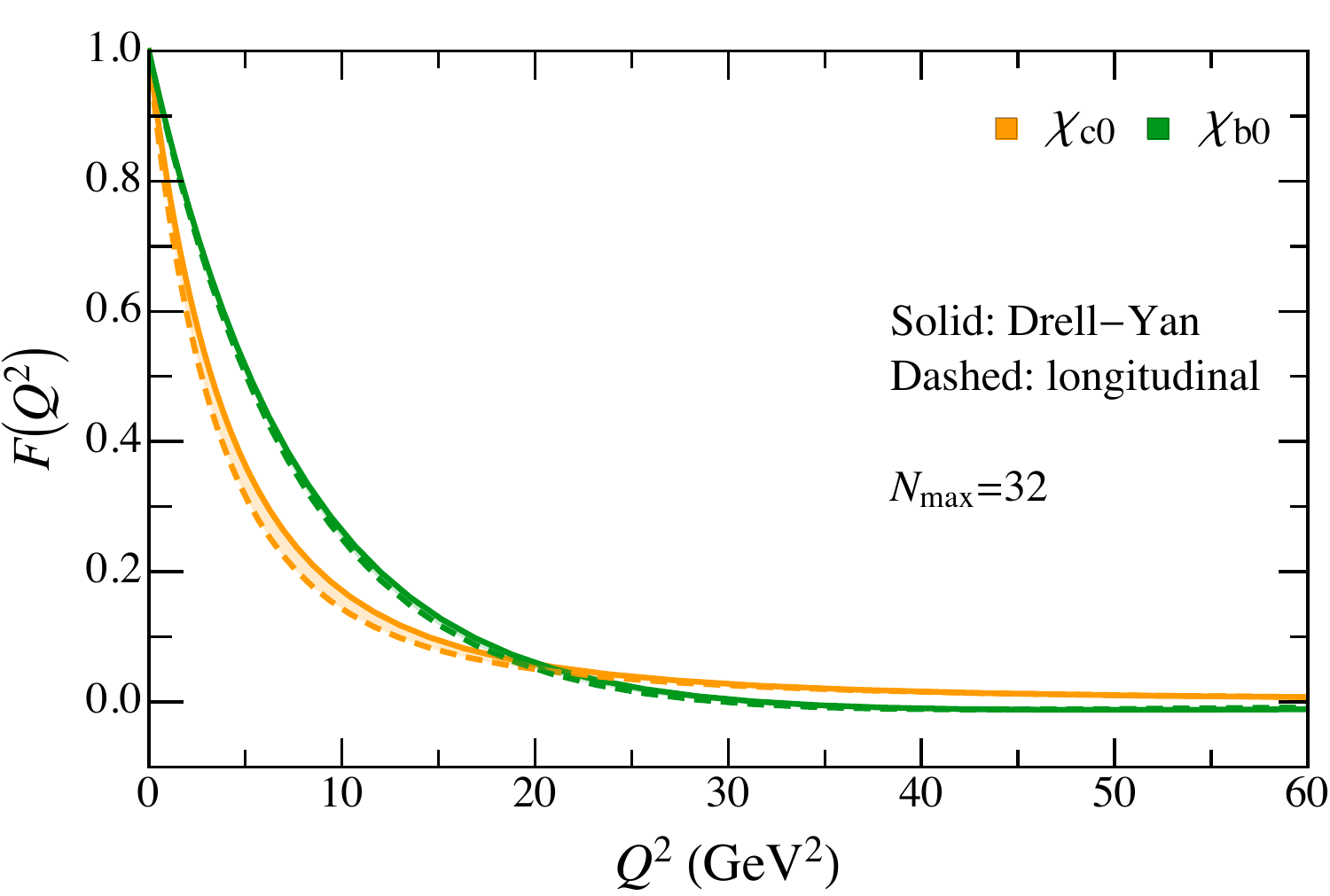}\\
\includegraphics[width=0.48\textwidth]{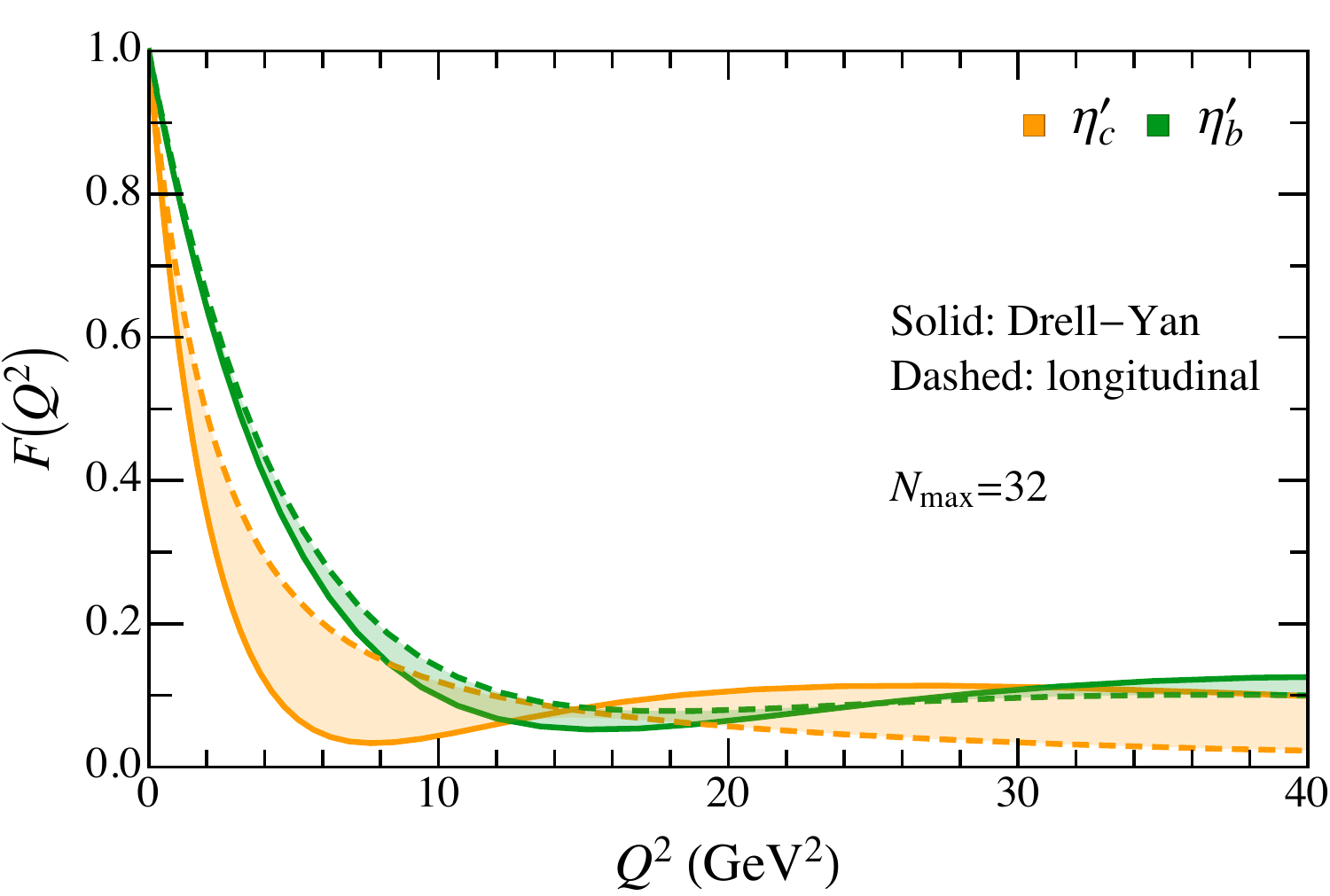} 
\caption{Frame dependence of heavy quarkonia form factors. 
The solid curves represent the Drell-Yan frame while the dashed curves 
the longitudinal frame. The shaded areas represent all other frames as obtained 
numerically through a dense sampling of the $z$ and 
$\Delta_{\perp}$ parameter space. Note that the appearance of crossing lines in the third panel of may be misleading since there is a spread in the distribution of dense points nearby that are not visible at the resolution of the figure. The basis is truncated with 
$N_{\max} = L_{\max} = 32$ (see text).}
\label{fig:heavy_quarkonia}
\end{figure}

Figure~\ref{fig:heavy_quarkonia} shows numerical results for charmonia $\eta_c$, $\chi_{c0}$, $\eta_c'$ and their bottomonium
counterparts $\eta_b$, $\chi_{b0}$, $\eta_b'$. 
A basis truncation $N_{\max} = L_{\max}=32$ is used. The solid curves represent the Drell-Yan frame while the dashed curves 
the longitudinal frame. The shaded areas represent all other frames as obtained numerically through a dense sampling of the $z$ and 
$\Delta_{\perp}$ parameter space. We observe that overall, the frame dependence is moderate
for both systems. The frame dependence of bottomonia is also smaller than that of charmonia, which is consistent 
with the fact that bottomonia are less relativistic.  These results also show
that the Drell-Yan frame and the longitudinal frame are indeed two special frames: their respective form factors typically signal the
extreme of the form factor $F(z, Q^2)$ for a given $Q^2$ and $0 \le z < 1$\footnote{This is most of the cases but not all.}.  
Therefore, the difference between the Drell-Yan frame and the longitudinal frame can be used to approximately characterize the frame dependence. 

\begin{figure}
\centering
\includegraphics[width=0.48\textwidth]{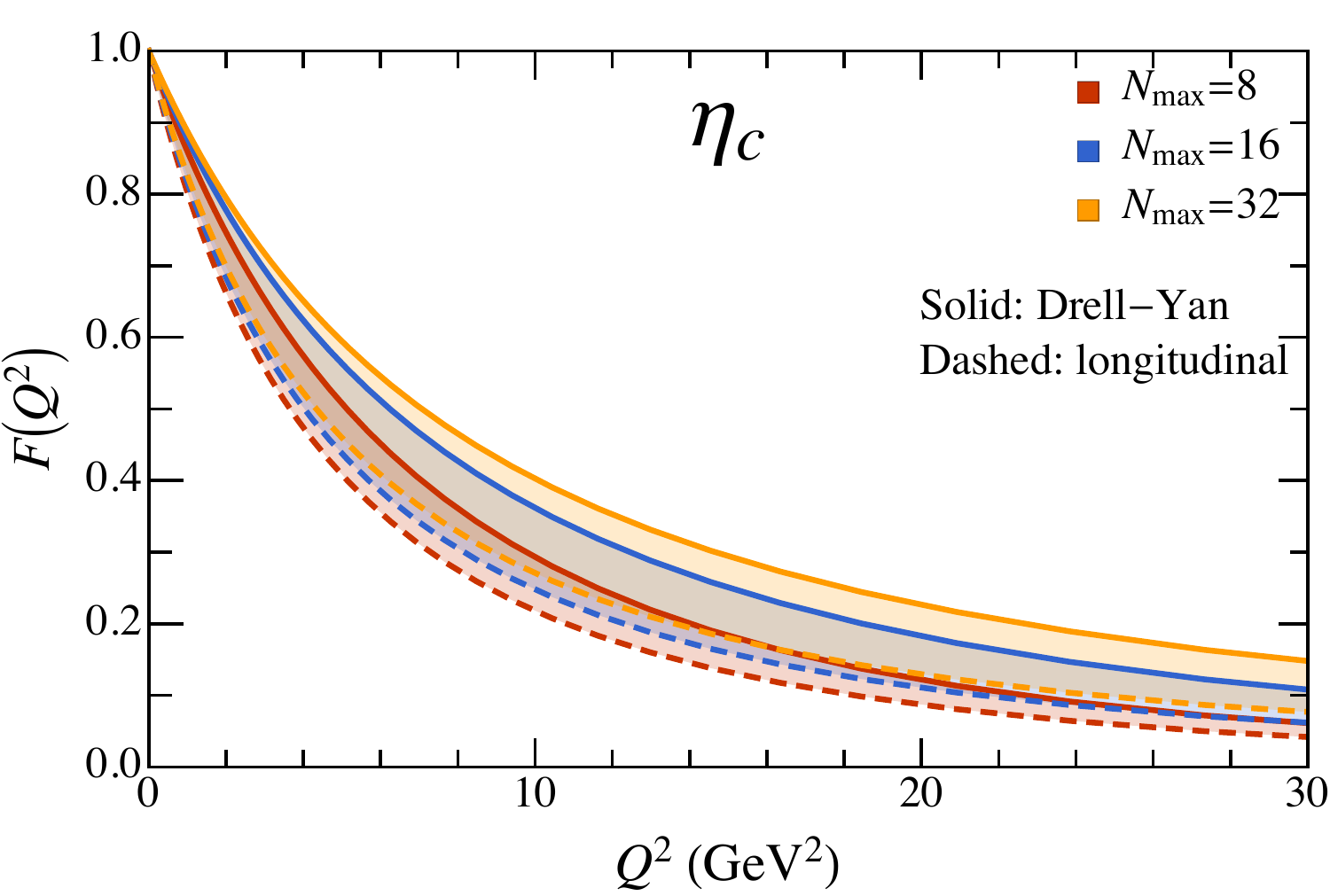}
\includegraphics[width=0.48\textwidth]{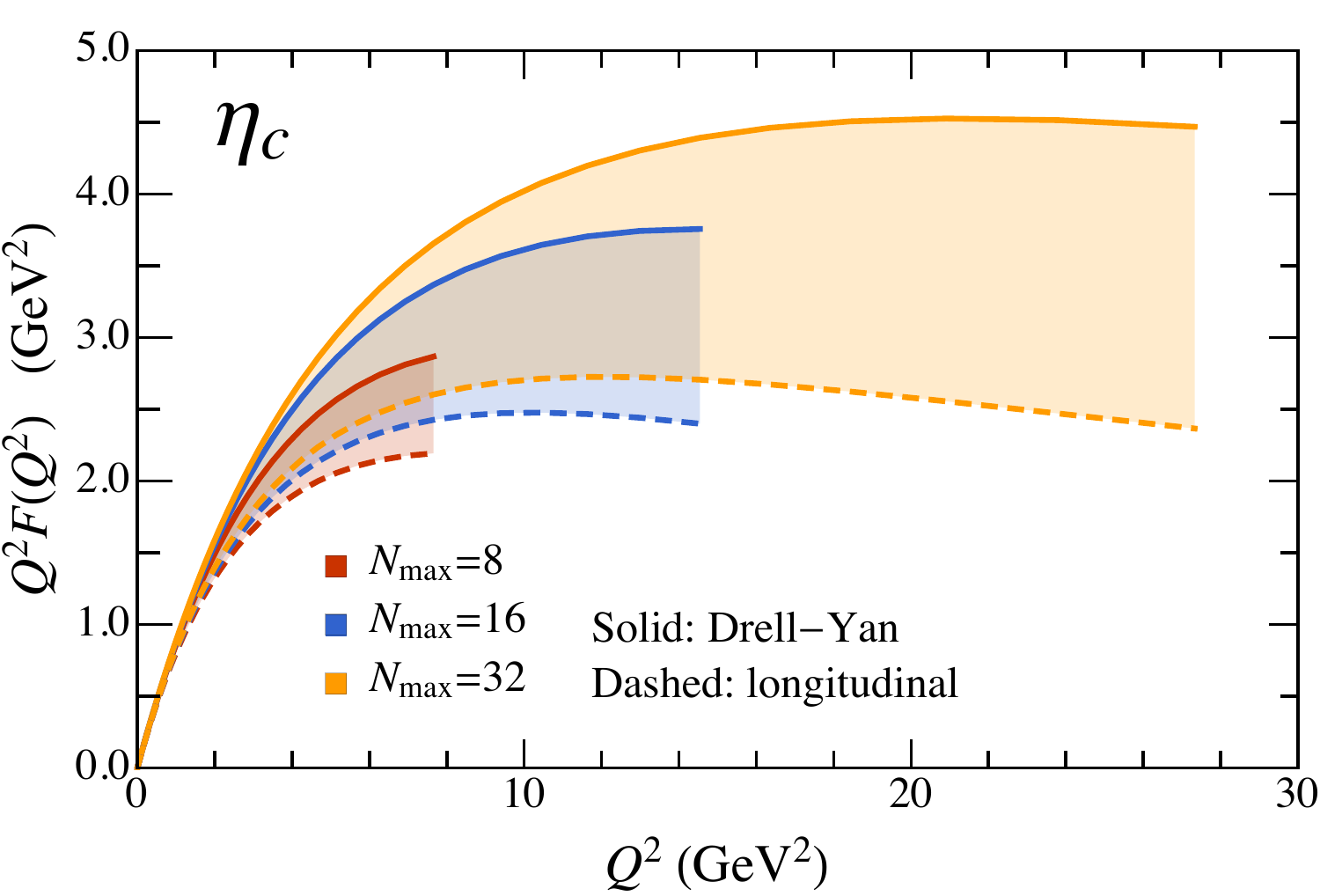}

\includegraphics[width=0.48\textwidth]{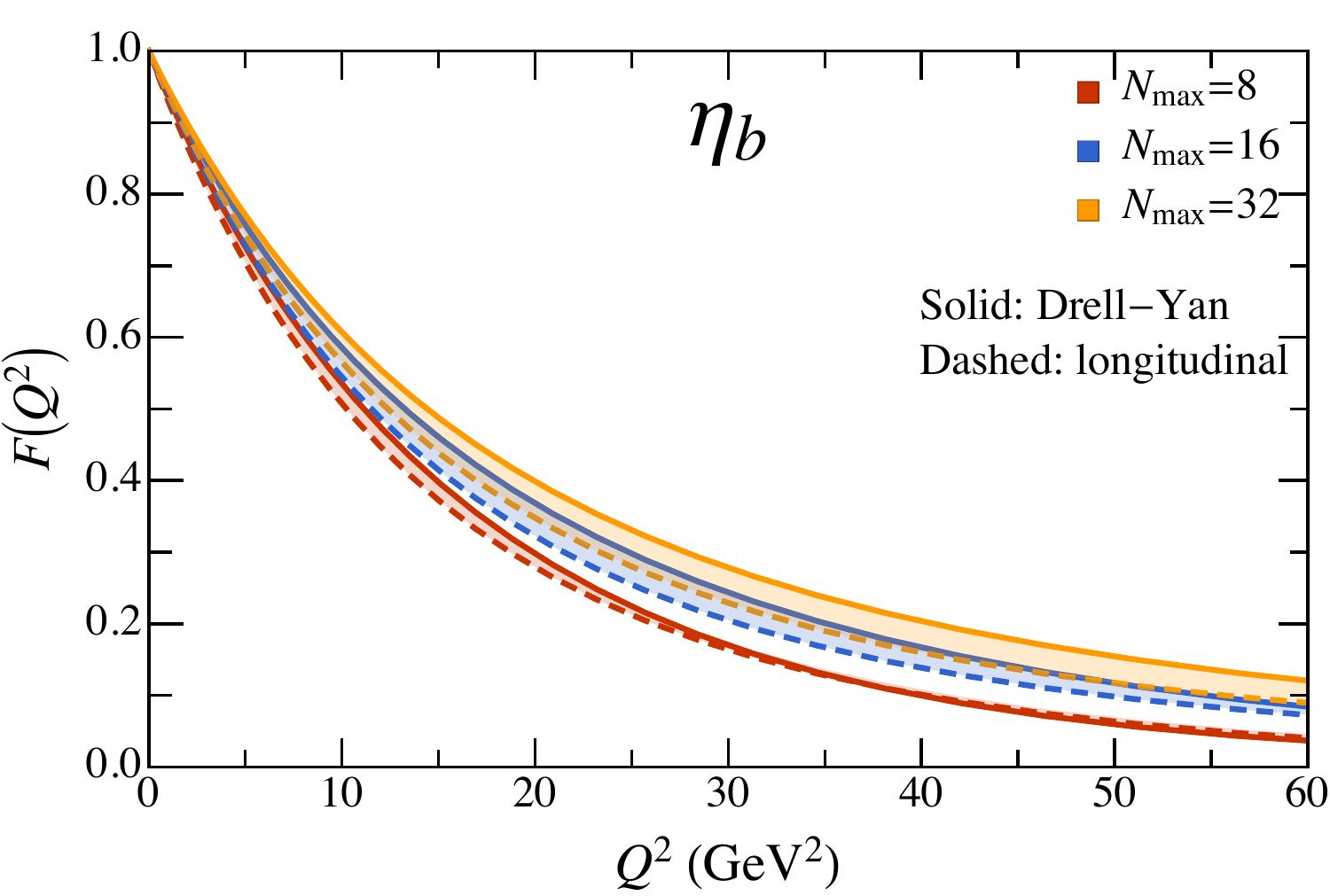}
\includegraphics[width=0.48\textwidth]{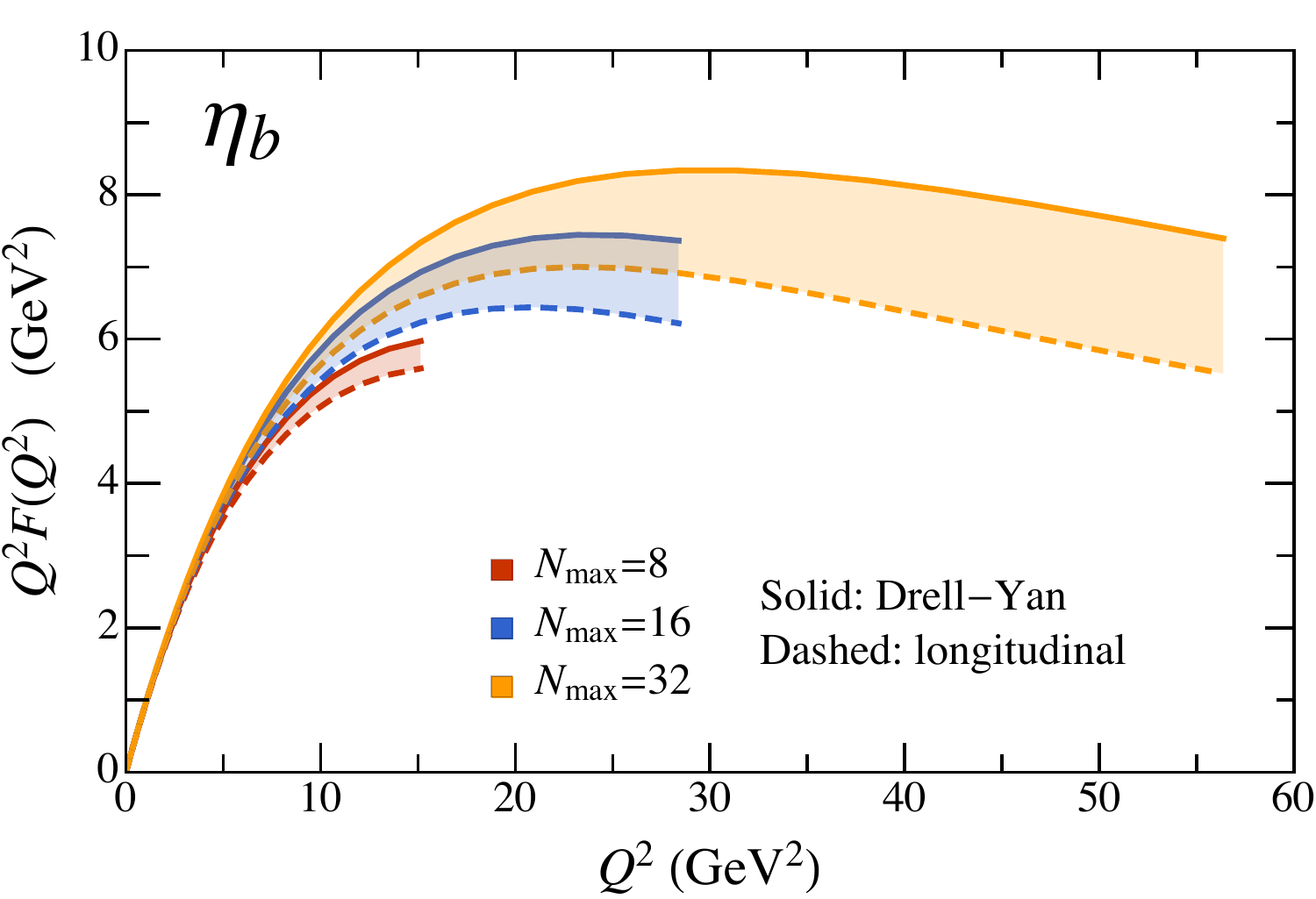}

\caption{Form factors with different basis truncations. On the right panels, form factors
are shown up to the UV scale $\Lambda_\textsc{uv} = \kappa\sqrt{N_{\max}}$. Note that
in the basis representation, this is a soft cutoff.}
\label{fig:basis}
\end{figure}

\begin{figure}
\centering
\includegraphics[width=0.48\textwidth]{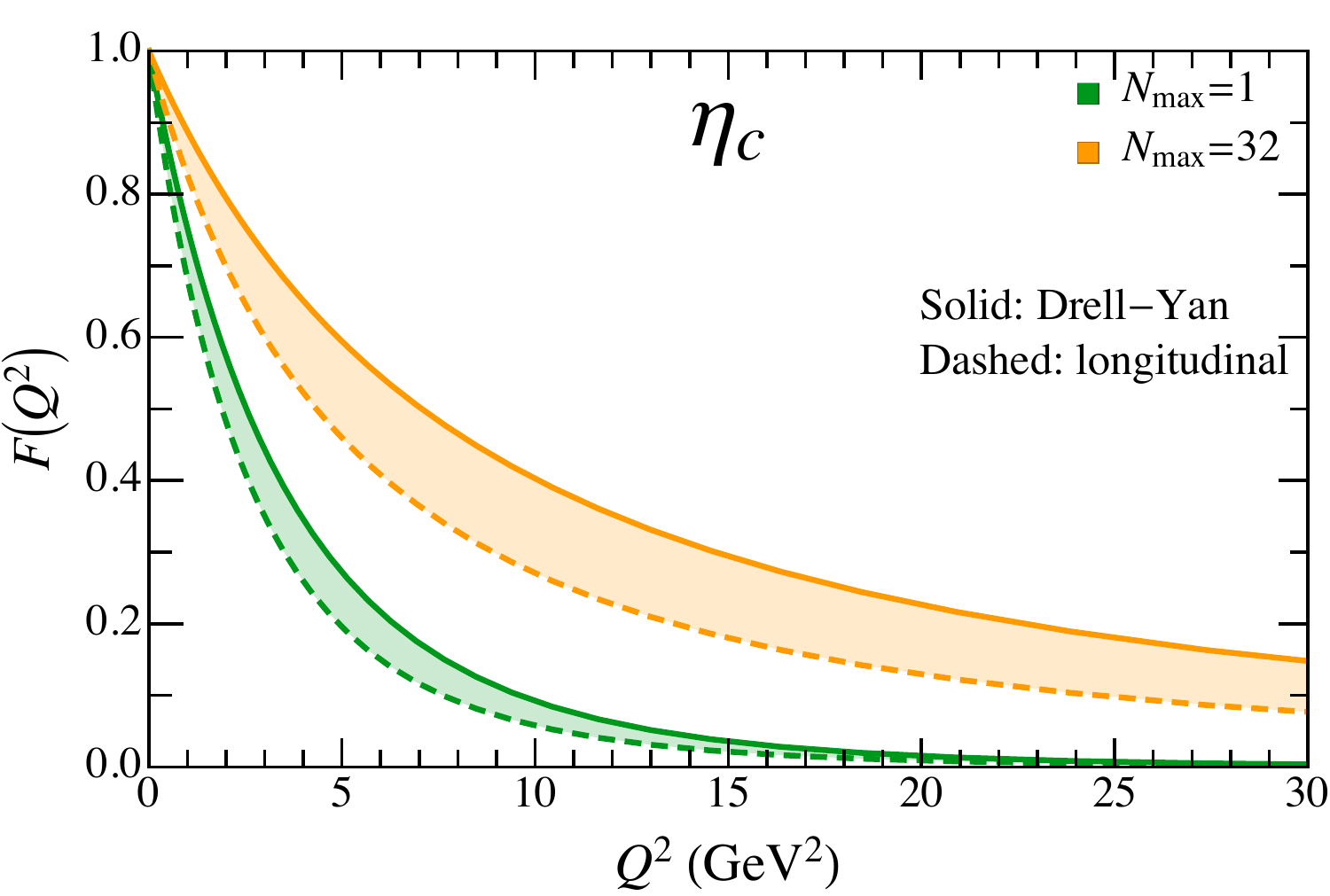}
\includegraphics[width=0.48\textwidth]{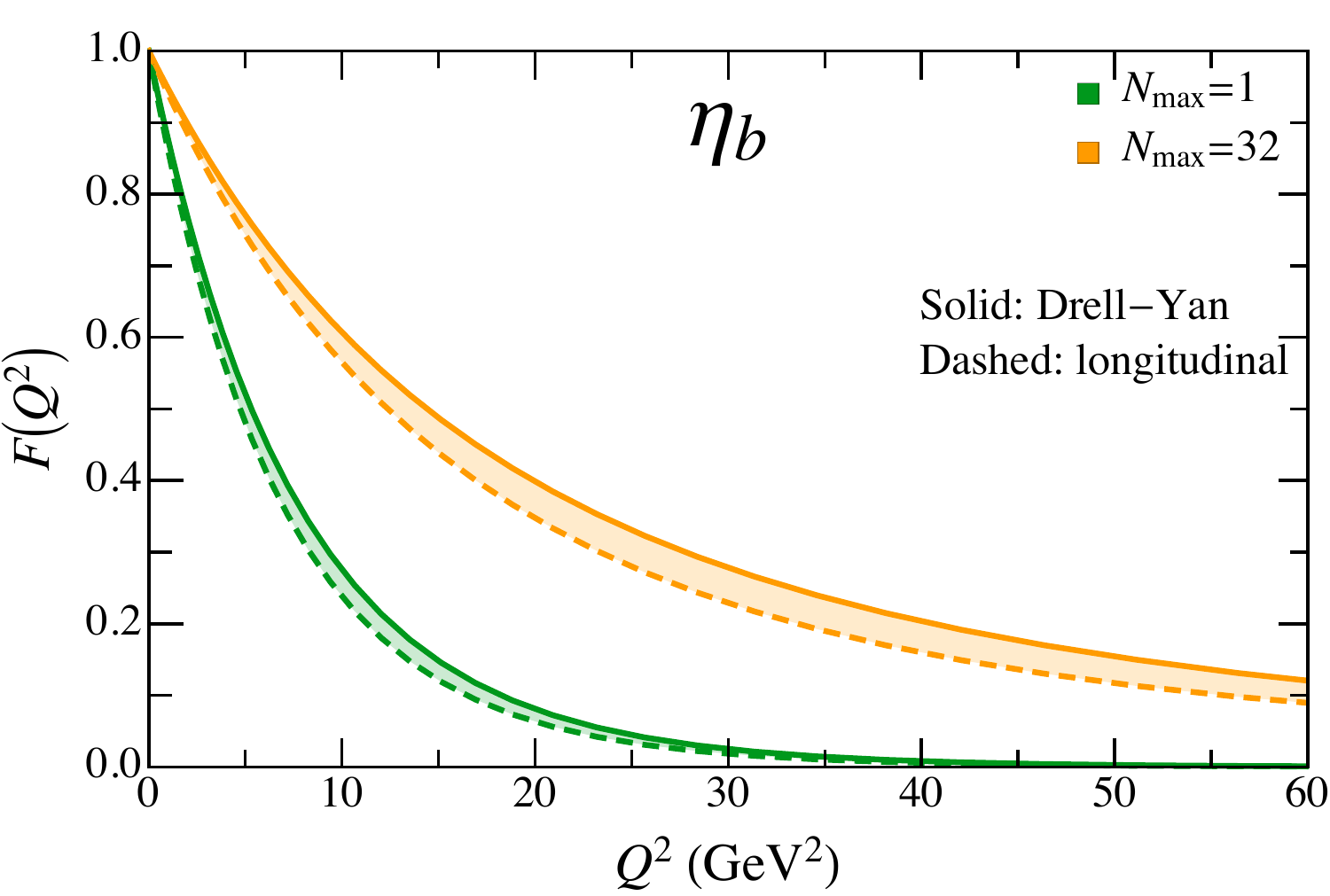}

\includegraphics[width=0.48\textwidth]{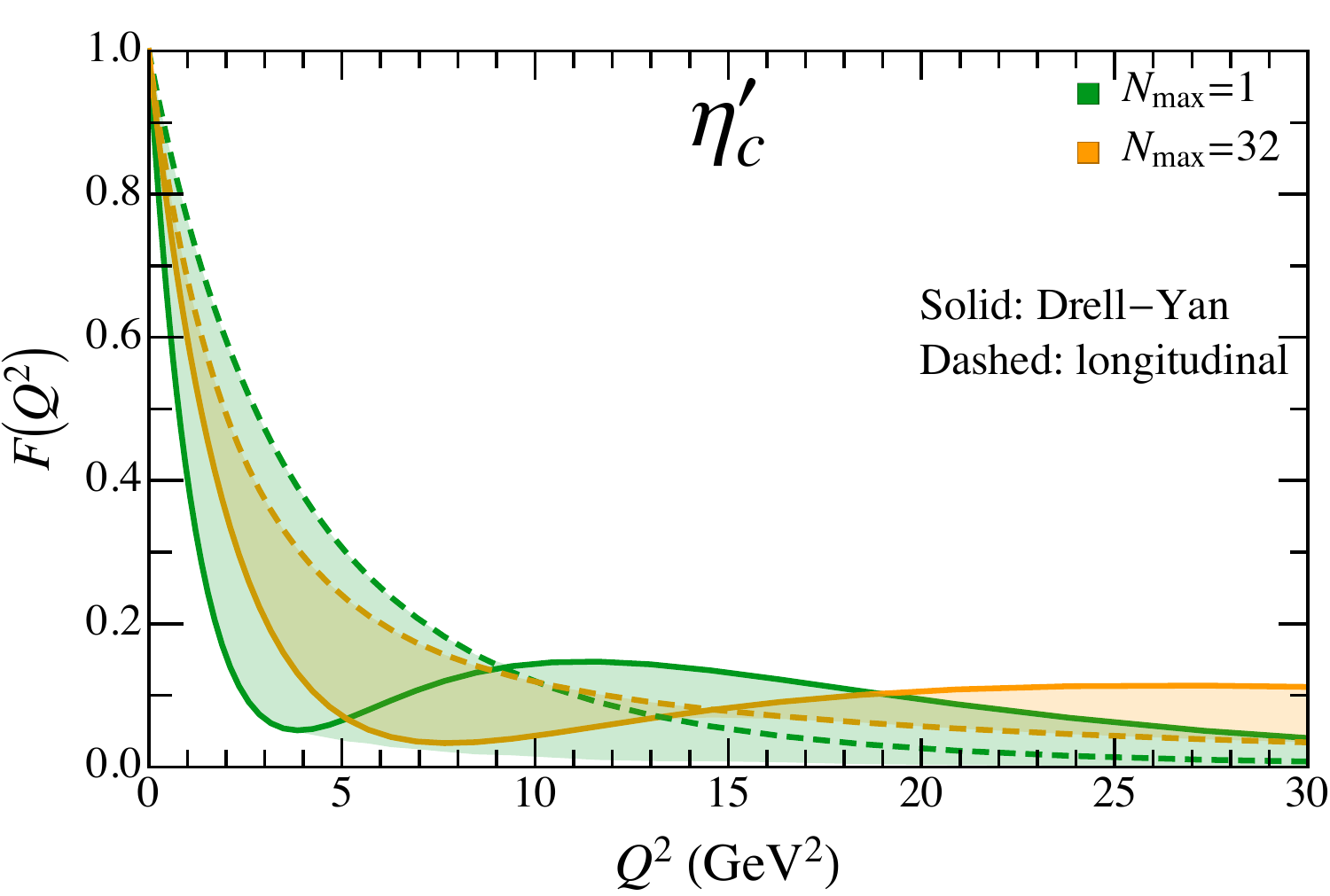}
\includegraphics[width=0.48\textwidth]{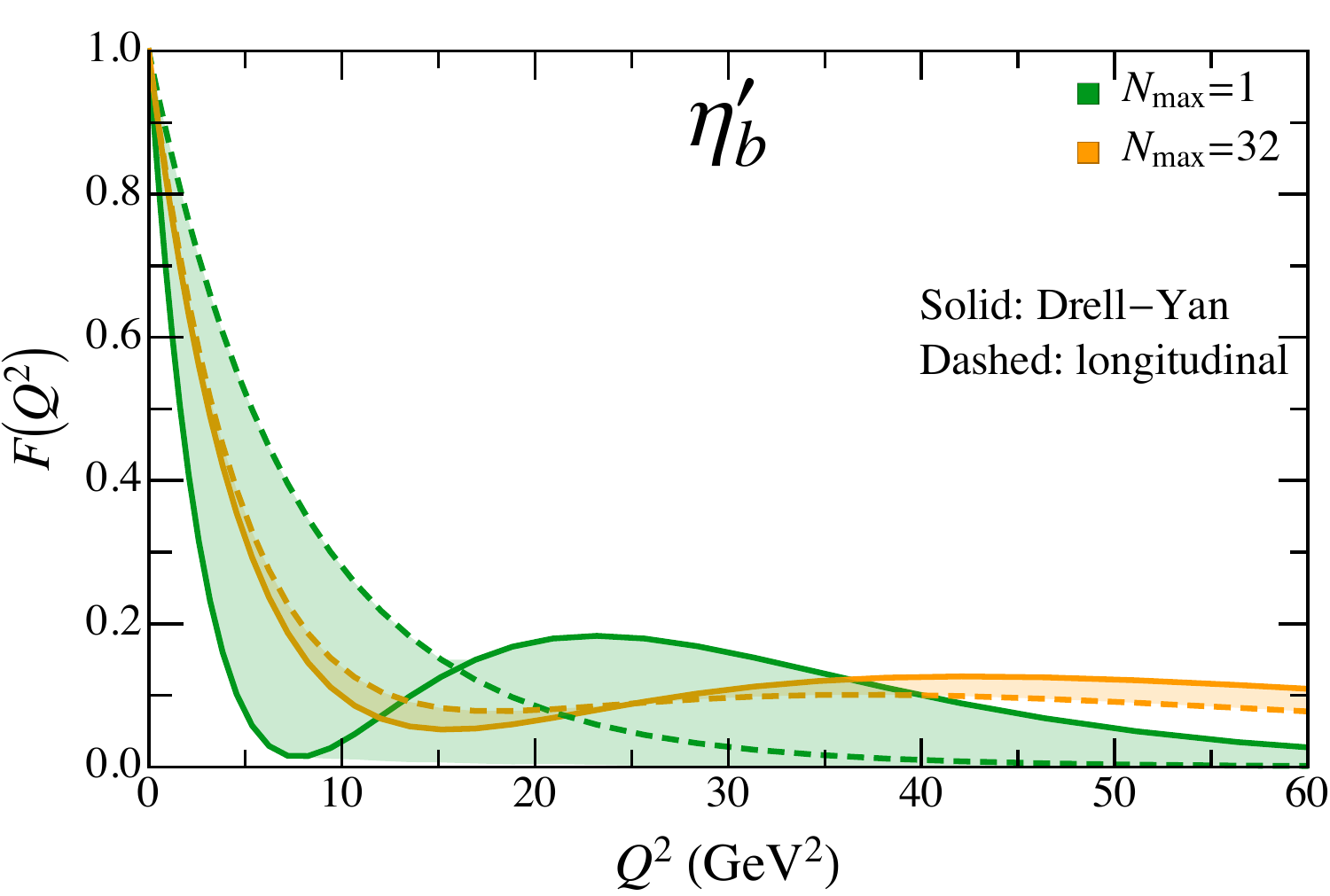}

\includegraphics[width=0.48\textwidth]{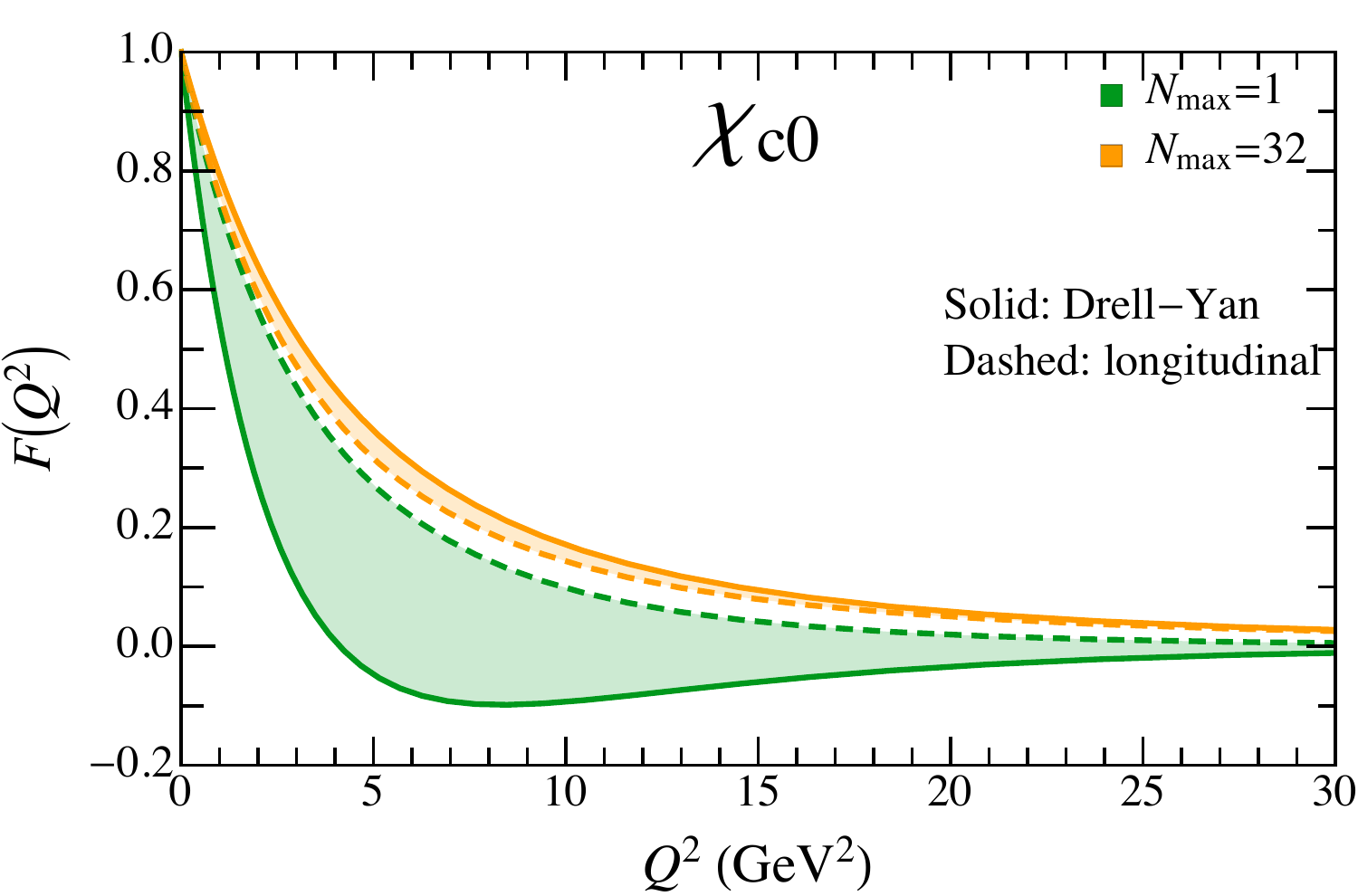}
\includegraphics[width=0.48\textwidth]{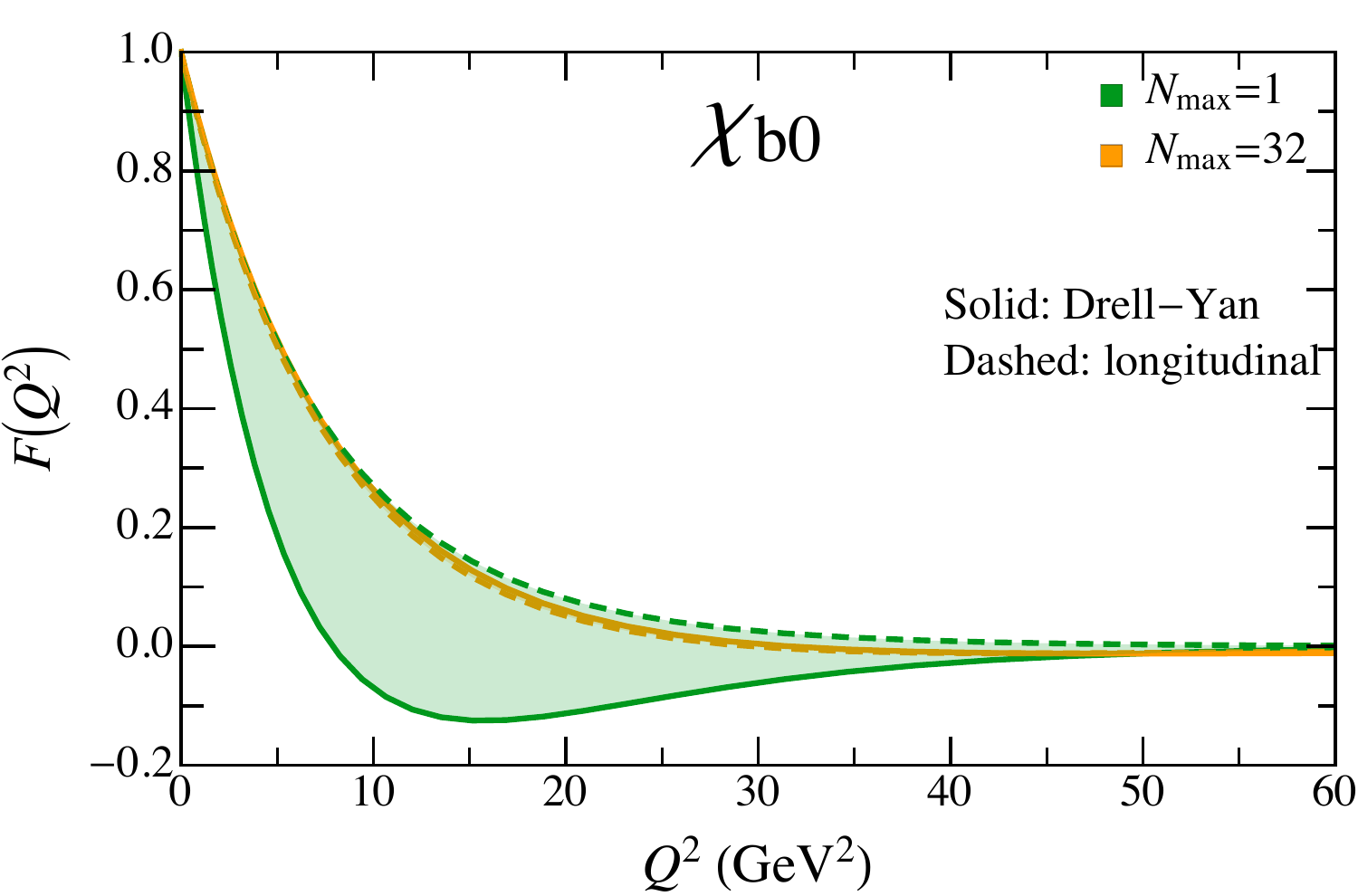}

\caption{Comparison of form factors from the leading basis function ($N_{\max}=L_{\max}=1$) and BLFQ ($N_{\max}=L_{\max}=32$).}
\label{fig:BLFH}
\end{figure}

In the LFWFs, there are two sources of Lorentz symmetry violation. One comes from the Fock sector truncation and the   associated effective interaction.  The other comes from the basis truncation. Form factors of $\eta_c$, $\eta_b$ from different basis truncations ($N_{\max} =L_{\max}=8,16,32$) are shown in Fig.~\ref{fig:basis}. The basis convergence 
is observed to be reasonable up to the UV limit specified by  the basis cutoff. 
 To further see the basis truncation effects, we compare the form factors
 evaluated from the leading basis function ($N_{\max}=L_{\max} = 1$) and the full diagonalization ($N_{\max}=L_{\max}=32$) in Fig.~\ref{fig:BLFH}.
 The basis function is the solution of the long-distance part, i.e. the light-front holographic
 QCD (LFHQCD), without the contributions from the one-gluon exchange interaction. Therefore, this is essentially a comparison between LFHQCD\footnote{In 
 the LFHQCD description of heavy quarkonia, a longitudinal function is required.  Ref.~\cite{Li:2017mlw} compares two popular
 longitudinal functions and found they are almost identical for heavy quarkonia. 
 } and BLFQ for a particular set of parameters. 
 We find that the excited states are more sensitive to the basis truncation. That is, we find that the excited states require more
 basis functions to resolve the spatial structure of the excited state wave functions. On the other hand, the ground states are more
 sensitive to the Fock sector truncation and to the model for the effective interaction. For example, the ground-state BLFQ form factors show more 
 frame dependence. This is because, compared  with LFHQCD (i.e. $N_{\max}=L_{\max}=1$), BLFQ has an additional interaction, the one-gluon exchange,
 which, upon Fock sector truncation, introduces an additional source of Lorentz symmetry violation. 
 
\section{Discussion and Conclusions}\label{sec:conclusion}

We have shown, using phenomenological LFWFs, scalar and pseudoscalar heavy quarkonia form factors 
admit moderate frame dependence. This dependence decreases from charmonium to bottomonium. We are therefore led to
to consider the frame dependence in light mesons. However, a similar model for light mesons is yet to be developed. As mentioned
above, the phenomenological model is based on LFHQCD, whose LFWFs can be readily used for light mesons.  
On the other hand, using similar phenomenological wave functions, Isgur and Smith found that the pion form factor has a large frame dependence \cite{Isgur:1988iw}. In Ref.~\cite{Sawicki:1992qj}, using a BSA, the discrepancy is attributed to the zero-mode contributions in the longitudinal frame. Such contributions are absent in the Drell-Yan frame. 

\begin{figure}
\centering 
\includegraphics[width=0.48\textwidth]{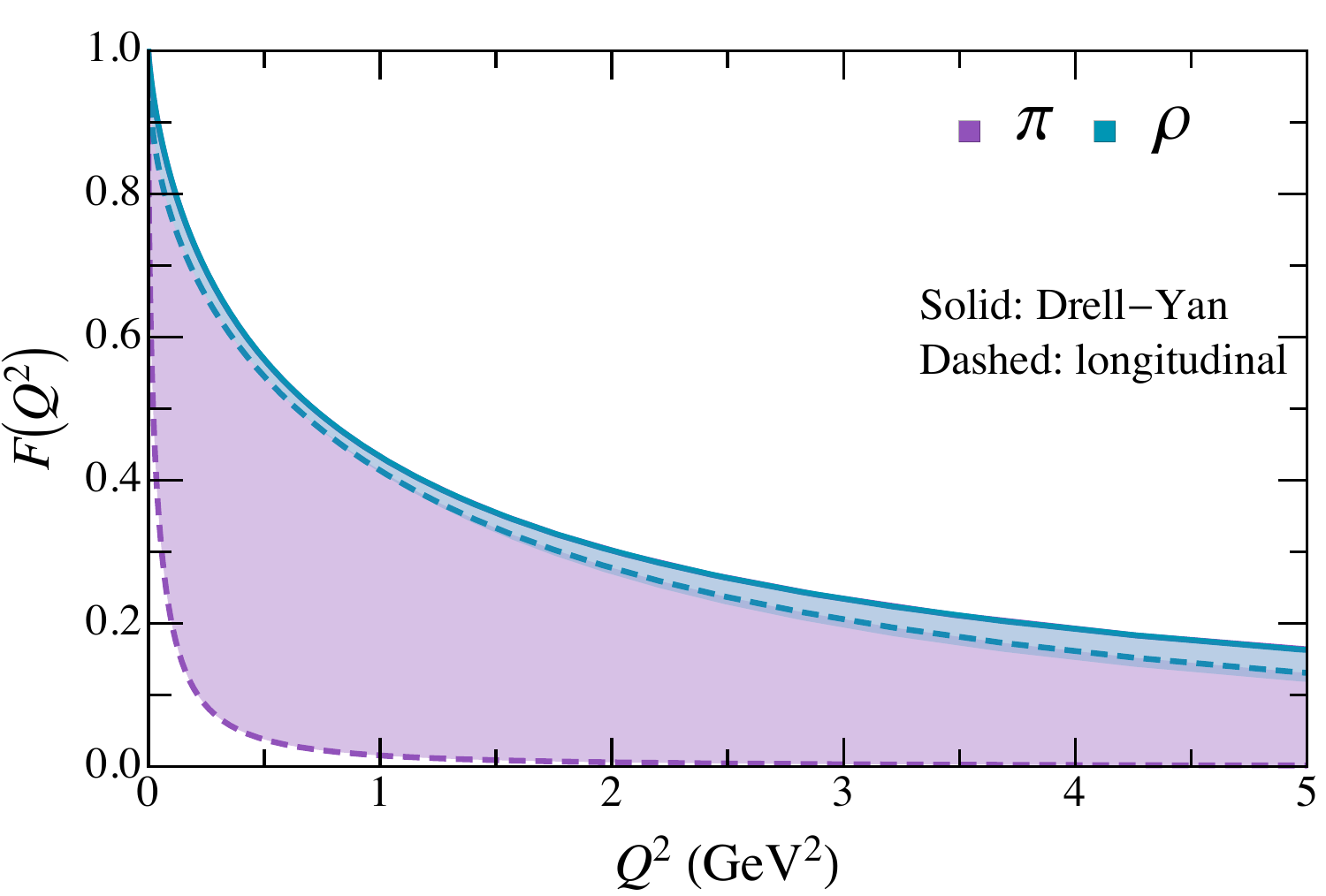}
\includegraphics[width=0.48\textwidth]{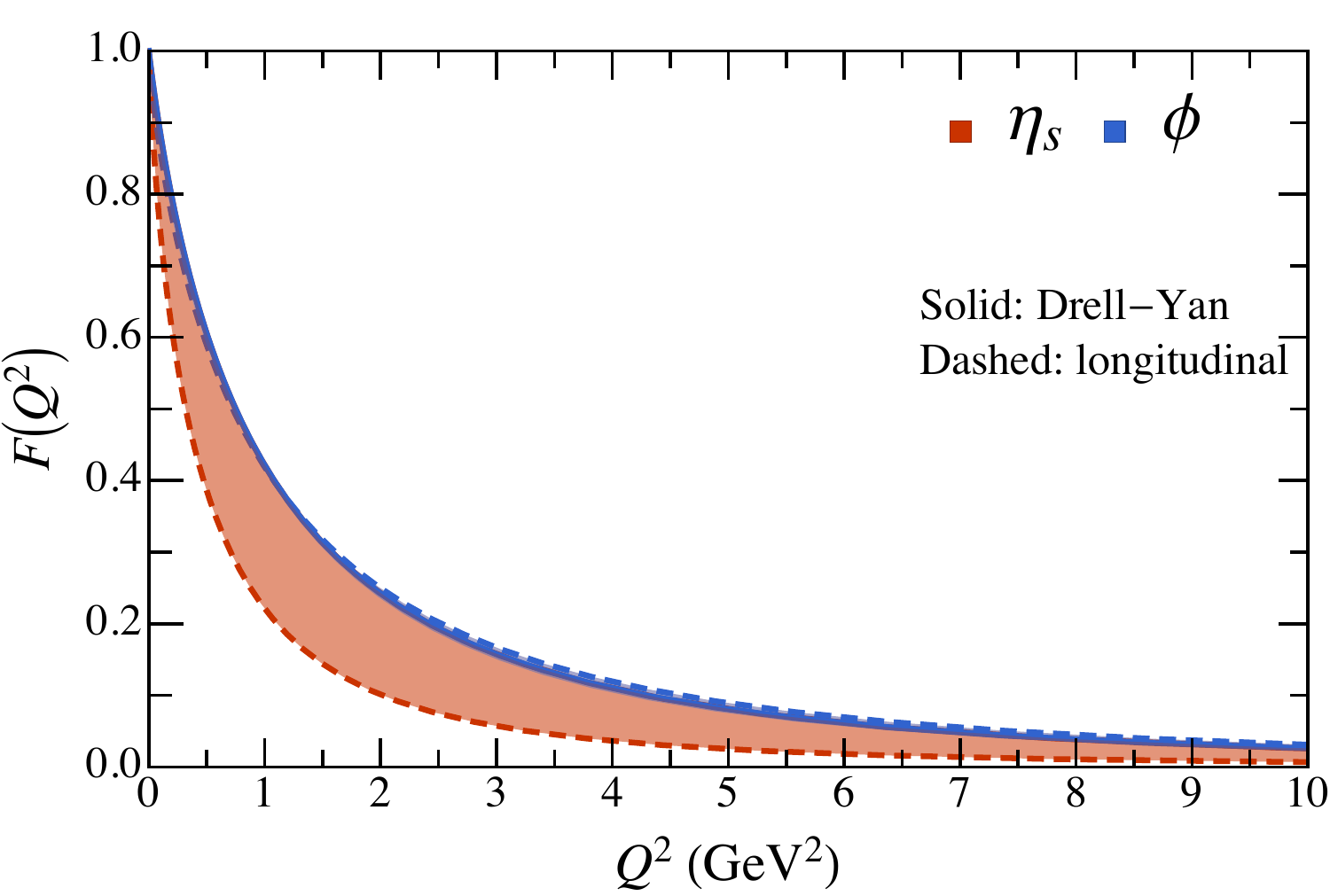}
\caption{Frame dependence of (left) $\pi$, $\rho$ and (right) $\eta_s$, $\phi$ form factors from LFHQCD. 
We adopt quark mass $m_q = 46\,\text{MeV}$, $m_s = 356\,\text{MeV}$ and confining scale
$\kappa = 0.54\,\text{GeV}$ \cite{Brodsky:2014yha}.
The solid curves represent
 form factors in the Drell-Yan frame, which are the same for $\pi$ and $\rho$ (left) as well as $\eta_s$ and $\phi$ (right) in LFHQCD. The dashed curves are
the form factors in the longitudinal frame. While both pseudoscalars display considerable frame dependence, that of $\pi$ 
seems especially large. }
\label{fig:pi}
\end{figure}

Figure~\ref{fig:pi} presents the $\pi$, $\rho$, $\eta_s$ and $\phi$ form factors obtained from LFHQCD\footnote{Here $\eta_s$ is the ground-state pseudoscalar $s\bar s$ meson, which does not have any correspondence in Nature. The physical pseudoscalars $\eta$ and $\eta'$ are dominated
by the axial anomaly, which is not described in LFHQCD.  $\eta_s$ is used as a theoretical construction in LFHQCD with a predicted mass $676$ MeV. }. 
Note that the spin effect is ignored in LFHQCD (cf. \cite{Leitner:2010nx, Ahmady:2016ufq, Yabusaki:2015dca, Mello:2016gvu, daSilva:2012gf}). So $\rho$ and $\pi$ share the same spatial LFWF hence form factors in the Drell-Yan frame\footnote{The longitudinal form factor depends
on the physical mass.}. Similarly, $\phi$ and $\eta_s$ share the same LFWF hence form factors in the Drell-Yan frame. 
These light vector mesons $\rho$, $\phi$ show relatively small frame dependence in charge form factors as compared to 
the pseudoscalar mesons. The frame dependence in pion is especially large. Comparing $\rho$ and $\pi$, this large frame dependence can be attributed to the mass, as this is the only difference between $\rho$ and $\pi$ in LFHQCD (spin is assigned, not dynamical). 

The mean-square radius $r_{\pi}^2$ controls the slope of the form factor at $Q^2\to0$. It can be shown
that, in the Drell-Yan frame, $r^2_\pi \propto \kappa^{-2} \ln M^2_{\pi}/\kappa^2 + O(M^2_\pi/\kappa^2)$, whereas in the longitudinal frame, $r^2_{\pi} \propto \kappa^2 / M^{4}_\pi+ O(M^2_\pi/\kappa^2)$
up to a log correction $\ln M_\pi/\kappa$. Here $\kappa$ is the confining scale parameter. As $M_\pi \ll \kappa$, the discrepancy in $r_\pi^2$ 
 explains the large frame dependence of the pion form factor at low $Q^2$. The asymptotics of $r_\pi^2$ obtained in the Drell-Yan frame is in agreement with the prediction
from chiral perturbation theory \cite{Beg:1973sc, Volkov:1974bi}. Therefore, this large frame dependence points to the (lack of) chiral symmetry breaking
in valence sector pion wave function, in particular, in the longitudinal direction. 
Chiral symmetry breaking on the light front is generally understood to require zero modes \cite{Wu:2003vn, Beane:2013ksa, Burkardt:1996pa, Burkardt:1998dd}. 
The conventional wisdom from the BSA suggest that this discrepancy may be caused by the (lack of) zero-mode contributions 
in the longitudinal frame, which have been omitted \cite{Sawicki:1992qj, Brodsky:1998hn}. Of course, without a corresponding BSA, it is not clear how to take such contributions
into account.
Therefore, developing a light-front model implementing dynamical chiral symmetry breaking is essential for a self-consistent description of 
the pion. 

To summarize, in this work, we develop a boost-invariant representation of space-like form factors in a general frame. 
We investigated the frame dependence of form factors in light-front dynamics. Heavy quarkonia are used as a 
concrete example. 
We show that the frame dependence is suppressed by the heavy quark mass. We identify the Drell-Yan frame and the
longitudinal frame as two special frames, whose difference can be used to represent the frame dependence.  

\section*{Acknowledgements}

The authors thank Alexander V. Smirnov, Stan Brodsky, Meijian Li, Christian Weiss, Raza Sufian and Tianbo Liu for valuable discussions. 
This work was supported in part by the US Department of
Energy (DOE) under Grant Nos.~DE-FG02-87ER40371, DE-FG02-04ER41302, DE-SC00{}18223 (SciDAC-4/NUCLEI) 
and DE-SC00{}15376 (DOE Topical Collaboration in Nuclear Theory for Double-Beta Decay 
and Fundamental Symmetries).
A portion of the computational resources were
provided by the National Energy Research Scientific Computing Center
(NERSC), which is supported by the US DOE Office of Science.

\appendix
\section{Light-front wave function representation beyond the valence sector}\label{sec:app}

In this section, we present the general LFWF representation of the elastic form factors
in a general frame with a boost-invariant parametrization using $z$ and $\vec\Delta_\perp$ ($0\le z \le 1$). The form factor admits 
a diagonal piece [Fig.~\ref{fig:LFWF_rep_a}] and an off-diagonal piece [Fig.~\ref{fig:LFWF_rep_b}]:
\begin{equation}
F(z, Q^2) = F_\text{diag}(z, Q^2) + F_\text{offdiag}(z, Q^2). 
\end{equation}
The diagonal part reads, 
\begin{equation}
F_\text{diag}(z, Q^2) = \frac{2}{2-z} \sum_n \int [\dd x_i \dd^2 k_{i\perp}]_n 
\sum_f q_f \sqrt{\frac{x_f}{x'_f}} 
 \psi_n^*(\{x_i, \vec k_{i\perp}, \lambda_i\}) 
 \psi_n(\{x'_i, \vec k'_{i\perp}, \lambda_i\}_f)
\end{equation}
where $x_f$ is the momentum fraction of the struck parton and $q_f$ its charge number. 
$\{x_i, \vec k_{i\perp}, \lambda_i\} = (x_1, \vec k_{1\perp}, \lambda_1, x_2, \vec k_{2\perp}, \lambda_2,
\cdots, x_n, \vec k_{n\perp}, \lambda_n)$ is a collection of parton quantum numbers for the $n$-body
Fock sector. 
$\{x_i', \vec k_{i\perp}', \lambda_i'\}_f$ is similar, except that a subscript $f$ is used to indicate the
dependence on the choice of struck parton. 
\begin{equation}
x'_i = \left\{
\begin{array}{ll}
x_i (1-z), & \; \text{ spectator}; \\
x_i (1-z) + z, & \; \text{ struck parton}.
\end{array}\right. 
\qquad
\vec k'_{i\perp} = \left\{
\begin{array}{ll}
\vec k_{i\perp} - x_i \vec \Delta_\perp, & \; \text{ spectator}; \\
\vec k_{i\perp} + (1-x_i)\vec\Delta_\perp, & \; \text{ struck parton}.
\end{array}\right.
\end{equation}
$[\dd x_i \dd^2 k_{i\perp}]_n$ is  the $n$-body phase space integration measure:
\begin{equation}
\int [\dd x_i \dd^2 k_{i\perp}]_n = S_n^{-1} \prod_{i=1}^n \sum_{\lambda_i} \int \frac{\dd x_i \dd^2k_{i\perp}}{2x_i (2\pi)^3}
2(2\pi)^3 \delta(\sum_i x_i -1 ) \delta^2(\sum_i \vec k_{i\perp}).
\end{equation}
In the Drell-Yan frame ($z=0$), the diagonal part reduces to the Drell-Yan-West formula \cite{Brodsky:1973kb, Brodsky:1980zm}.

The off-diagonal part reads, 
\begin{multline}
F_\text{offdiag}(z, Q^2) = \frac{2}{(2-z)} \sum_n \int [\dd x_i \dd^2 k_{i\perp}]_n 
\sum_f q_f 
\sum_{\lambda_f} \int_0^1\frac{\dd\xi}{2\xi(1-\xi)}\int\frac{\dd^2k_\perp}{(2\pi)^3} 
\\
\times \sqrt{\xi(1-\xi)}  \psi_n^*(\{x_i, \vec k_{i\perp}, \lambda_i\}) 
 \psi_{n+2}(\{x'_i, \vec k'_{i\perp}, \lambda_i'\}_f)
\end{multline}
where $x_f$ is the momentum fraction of the struck parton and $q_f$ its charge number, 
$[\dd x_i \dd^2 k_{i\perp}]_n$ is  again the $n$-body phase space integration measure.
The collections of parton quantum numbers $\{x_i, \vec k_{i\perp}, \lambda_i\}$ and
$\{x'_i, \vec k'_{i\perp}, \lambda'_i\}_f$ are similar to the diagonal part, except that, now, it is
understood the spectators within the initial and final states pair up.
\begin{equation}
x'_i = \left\{
\begin{array}{ll}
x_i (1-z), & \; \text{ spectator}; \\
z\xi, & \; \text{ struck quark}; \\
z(1-\xi), & \; \text{ struck antiquark}.
\end{array}\right. 
\qquad
\vec k'_{i\perp} = \left\{
\begin{array}{ll}
\vec k_{i\perp} - x_i \vec \Delta_\perp, & \; \text{ spectator}; \\
\vec k_{\perp} + \xi \vec\Delta_\perp, & \; \text{ struck quark}; \\
-\vec k_{\perp} +(1-\xi) \vec\Delta_\perp, & \; \text{ struck antiquark}.
\end{array}\right. 
\end{equation}
\begin{equation}
\lambda'_i = \left\{
\begin{array}{ll}
\lambda_i, & \; \text{ spectator}; \\
\lambda_f, & \; \text{ struck quark}; \\
-\lambda_f, & \; \text{ struck antiquark}.
\end{array}\right.
\end{equation}
In the Drell-Yan frame ($z=0$), the off-diagonal part has only zero-mode contributions. 

%%%%%%%%%%%%%%%%%%%%%%%%%%%%%%%%%%%%%%%%%%%%%%%%%%%

\end{document}